\documentclass{JINST}

\usepackage{graphicx}

\newcommand{\lsim}{\mathrel{\mathop{\kern 0pt \rlap
  {\raise.2ex\hbox{$<$}}}
  \lower.9ex\hbox{\kern-.190em $\sim$}}}
\newcommand{\gsim}{\mathrel{\mathop{\kern 0pt \rlap
  {\raise.2ex\hbox{$>$}}}
  \lower.9ex\hbox{\kern-.190em $\sim$}}}

\newcommand{\alt}{\mathrel{\mathop{\kern 0pt \rlap
  {\raise.2ex\hbox{$<$}}}
  \lower.9ex\hbox{\kern-.190em $\sim$}}}
\newcommand{\agt}{\mathrel{\mathop{\kern 0pt \rlap
  {\raise.2ex\hbox{$>$}}}
  \lower.9ex\hbox{\kern-.190em $\sim$}}}

\newcommand{\gagamma}{g_{a\gamma}}
\newcommand{\ckcs}{counts keV$^{-1}$~cm$^{-2}$~s$^{-1}$ }

\title{Conceptual Design of the International Axion Observatory (IAXO)}

\newcommand{\IRFU}{$^{1}$}
\newcommand{\SCarolina}{$^{2}$}
\newcommand{\CERN}{$^{3}$}
\newcommand{\SPTSaclay}{$^{4}$}
\newcommand{\INFN}{$^{5}$}
\newcommand{\UNIZAR}{$^{6}$}
\newcommand{\LLNL}{$^{7}$}
\newcommand{\LBNL}{$^{8}$}
\newcommand{\DogusU}{$^{9}$}
\newcommand{\UHaifa}{$^{10}$}
\newcommand{\DTUSpace}{$^{11}$}
\newcommand{\NPI}{$^{12}$}
\newcommand{\UBonn}{$^{13}$}
\newcommand{\DESY}{$^{14}$}
\newcommand{\ArisU}{$^{15}$}
\newcommand{\NCSR}{$^{16}$}
\newcommand{\UVEG}{$^{17}$}
\newcommand{\INR}{$^{18}$}
\newcommand{\BGU}{$^{19}$}
\newcommand{\Columbia}{$^{20}$}
\newcommand{\Kyoto}{$^{21}$}
\newcommand{\Darmstadt}{$^{22}$}
\newcommand{\Hamburg}{$^{23}$}
\newcommand{\ICE}{$^{24}$}
\newcommand{\JAEA}{$^{25}$}
\newcommand{\UHEI}{$^{26}$}
\newcommand{\RBI}{$^{27}$}
\newcommand{\TokyoU}{$^{28}$}
\newcommand{\URijeka}{$^{29}$}
\newcommand{\MatKyoto}{$^{30}$}
\newcommand{\MPI}{$^{31}$}
\newcommand{\TokyoIT}{$^{32}$}
\newcommand{\BNL}{$^{33}$}
\newcommand{\UF}{$^{34}$}
\newcommand{\Berkeley}{$^{35}$}
\newcommand{\CapeTown}{$^{36}$}
\newcommand{\FNAL}{$^{37}$}
\newcommand{\PatrasU}{$^{38}$}

\newcommand{\SCarolinaName}{Physics Department, University of South Carolina, Columbia, SC, USA}
\newcommand{\SPTSaclayName}{IPHT, Centre d'\'Etudes de Saclay (CEA-Saclay), Gif-sur-Yvette, France}
\newcommand{\INFNName}{Instituto Nazionale di Fisica Nucleare (INFN), Sezione di Trieste and Universit\`a di Trieste, Trieste, Italy}
\newcommand{\UNIZARName}{Laboratorio de F\'{\i}sica Nuclear y Altas Energ\'{\i}as, Universidad de Zaragoza, Zaragoza, Spain}
\newcommand{\CERNName}{European Organization for Nuclear Research (CERN), Gen\`eve, Switzerland}
\newcommand{\LBNLName}{Lawrence Berkeley National Laboratory, USA}
\newcommand{\DogusUName}{Dogus University, Istanbul, Turkey}
\newcommand{\DTUSpaceName}{Technical University of Denmark, DTU Space Kgs. Lyngby, Denmark}
\newcommand{\UHaifaName}{Physics Department, University of Haifa, Haifa, 31905 Israel}
\newcommand{\IRFUName}{CEA Irfu, Centre de Saclay, F-91191 Gif-sur-Yvette, France}
\newcommand{\NPIName}{St. Petersburg Nuclear Physics Institute, St. Petersburg, Russia}
\newcommand{\UBonnName}{Physikalisches Institut der Universit\"{a}t Bonn, Bonn, Germany}
\newcommand{\ArisUName}{Aristotle University of Thessaloniki, Thessaloniki, Greece}
\newcommand{\NCSRName}{National Center for Scientific Research ``Demokritos'', Athens, Greece}
\newcommand{\UVEGName}{Instituto de Ciencias de las Materiales, Universidad de Valencia, Valencia, Spain}
\newcommand{\INRName}{Institute for Nuclear Research (INR), Russian Academy of Sciences, Moscow, Russia}
\newcommand{\URijekaName}{University of Rijeka, Croatia}
\newcommand{\BGUName}{Physics department, Ben Gurion Uiversity, Beer Sheva, Israel}
\newcommand{\ColumbiaName}{Columbia Astrophysics Laboratory, New York, USA}
\newcommand{\KyotoName}{Yukawa Institute for Theoretical Physics, Kyoto University, Kyoto, Japan}
\newcommand{\DarmstadtName}{Technische Universit\"{a}t Darmstadt, IKP, Darmstadt, Germany}
\newcommand{\HamburgName}{Institut f\"{u}r Experimentalphysik, Universit\"{a}t Hamburg, 22761 Hamburg, Germany}
\newcommand{\ICEName}{Institut de Ci\`encies de l'Espai (CSIC-IEEC), Facultat de Ci\`encies, Campus UAB, Bellaterra, Spain}
\newcommand{\JAEAName}{Advanced Science Research Center, Japan Atomic Energy Agency, Tokai-mura, Ibaraki-ken, Japan}
\newcommand{\UHEIName}{Institut f\"ur theoretische Physik, Universit\"at Heidelberg, Philosophenweg 16, 69120 Heidelberg, Germany}
\newcommand{\RBIName}{Rudjer Bo\v{s}kovi\'{c} Institute, Zagreb, Croatia}
\newcommand{\MatKyotoName}{Research Center for Low Temperature and Materials Sciences, Kyoto University, Kyoto, 606-8502 Japan}
\newcommand{\TokyoUName}{Institute for Cosmic Ray Research, University of Tokyo, Tokyo, Japan}
\newcommand{\TokyoITName}{Department of Physics, Tokyo Institute of Technology, Tokyo, Japan}
\newcommand{\DESYName}{Deutsches Elektronen-Synchrotron DESY, Hamburg, Germany}
\newcommand{\LLNLName}{Lawrence Livermore National Laboratory, Livermore, CA, USA}
\newcommand{\MPIName}{Max-Planck-Institut f\"{u}r Physik, Munich, Germany}
\newcommand{\BNLName}{Physics Department, Brookhaven National Lab, Upton, NY, USA}
\newcommand{\UFName}{Department of Physics, University of Florida, Gainesville, FL 32611, USA}
\newcommand{\BerkeleyName}{Department of Nuclear Engineering, University of California Berkeley, Berkeley, CA, USA}
\newcommand{\CapeTownName}{University of Cape Town, South Africa}
\newcommand{\FNALName}{Fermi National Accelerator Laboratory, Batavia, IL, USA}
\newcommand{\PatrasUName}{Physics Department, University of Patras, Patras, Greece}

\newcommand{\IAXOAuthorList}{

E.~Armengaud\IRFU,
F.~T.~Avignone\SCarolina,
M.~Betz\CERN,
P.~Brax\SPTSaclay,
P.~Brun\IRFU,
G.~Cantatore\INFN,
J.~M.~Carmona\UNIZAR,
G.~P.~Carosi\LLNL,
F.~Caspers\CERN,
S.~Caspi\LBNL,
S.~A.~Cetin\DogusU,
D.~Chelouche\UHaifa,
F.~E.~Christensen\DTUSpace,
A.~Dael\IRFU,
T.~Dafni\UNIZAR,
M.~Davenport\CERN,
A.V.~Derbin\NPI,
K.~Desch\UBonn,
A.~Diago\UNIZAR,
B.~D\"obrich\DESY,
I.~Dratchnev\NPI,
A.~Dudarev\CERN,
C.~Eleftheriadis\ArisU,
G.~Fanourakis\NCSR,
E.~Ferrer-Ribas\IRFU,
J.~Gal\'an\IRFU,
J.~A.~Garc\'ia\UNIZAR,
J.~G.~Garza\UNIZAR,
T.~Geralis\NCSR,
B.~Gimeno\UVEG,
I.~Giomataris\IRFU,
S.~Gninenko\INR,
H.~G\'omez\UNIZAR,
D.~Gonz\'{a}lez-D\'{\i}az\UNIZAR,
E.~Guendelman\BGU,
C.~J.~Hailey\Columbia,
T.~Hiramatsu\Kyoto,
D.~H.~H.~Hoffmann\Darmstadt,
D.~Horns\Hamburg,
F.~J.~Iguaz\UNIZAR,
I.~G.~Irastorza\UNIZAR$^{,}$\footnote{Spokesperson. E-mail: irastorz@unizar.es},
J.~Isern\ICE,
K.~Imai\JAEA,
A.~C.~Jakobsen\DTUSpace,
J.~Jaeckel\UHEI,
K.~Jakov\v{c}i\'{c}\RBI,
J.~Kaminski\UBonn,
M.~Kawasaki\TokyoU,
M.~Karuza\URijeka,
M.~Kr\v{c}mar\RBI,
K.~Kousouris\CERN,
C.~Krieger\UBonn,
B.~Laki\'{c}\RBI,
O.~Limousin\IRFU,
A.~Lindner\DESY,
A.~Liolios\ArisU,
G.~Luz\'on\UNIZAR,
S.~Matsuki\MatKyoto,
V.~N.~Muratova\NPI,
C.~Nones\IRFU,
I.~Ortega\UNIZAR,
T.~Papaevangelou\IRFU,
M.~J.~Pivovaroff\LLNL,
G.~Raffelt\MPI,
J.~Redondo\MPI,
A.~Ringwald\DESY,
S.~Russenschuck\CERN,
J.~Ruz\LLNL,
K.~Saikawa\TokyoIT,
I.~Savvidis\ArisU,
T.~Sekiguchi\TokyoU,
Y.~K.~Semertzidis\BNL,
I.~Shilon\CERN,
P.~Sikivie\UF,
H.~Silva\CERN,
H.~ten~Kate\CERN,
A.~Tomas\UNIZAR,
S.~Troitsky\INR,
T.~Vafeiadis\CERN,
K.~van~Bibber\Berkeley,
P.~Vedrine\IRFU,
J.~A.~Villar\UNIZAR,
J.~K.~Vogel\LLNL,
L. Walckiers\CERN,
A.~Weltman\CapeTown,
W.~Wester\FNAL,
S.~C.~Yildiz\DogusU,
K.~Zioutas\PatrasU

}

\newcommand{\IAXOAffiliationList}{
\noindent
\IRFU\IRFUName\\
\SCarolina\SCarolinaName\\
\CERN\CERNName\\
\SPTSaclay\SPTSaclayName\\
\INFN\INFNName\\
\UNIZAR\UNIZARName\\
\LLNL\LLNLName\\
\LBNL\LBNLName\\
\DogusU\DogusUName\\
\UHaifa\UHaifaName\\
\DTUSpace\DTUSpaceName\\
\NPI\NPIName\\
\UBonn\UBonnName\\
\DESY\DESYName\\
\ArisU\ArisUName\\
\NCSR\NCSRName\\
\UVEG\UVEGName\\
\INR\INRName\\
\BGU\BGUName\\
\Columbia\ColumbiaName\\
\Kyoto\KyotoName\\
\Darmstadt\DarmstadtName\\
\Hamburg\HamburgName\\
\ICE\ICEName\\
\JAEA\JAEAName\\
\UHEI\UHEIName\\
\RBI\RBIName\\
\TokyoU\TokyoUName\\
\URijeka\URijekaName\\
\MatKyoto\MatKyotoName\\
\MPI\MPIName\\
\TokyoIT\TokyoITName\\
\BNL\BNLName\\
\UF\UFName\\
\Berkeley\BerkeleyName\\
\CapeTown\CapeTownName\\
\FNAL\FNALName\\
\PatrasU\PatrasUName\\
}

\author{\IAXOAuthorList
\\
\IAXOAffiliationList

E-mail: \email{Igor.Irastorza@cern.ch}

}

\abstract{The International Axion Observatory (IAXO) will be a forth generation axion helioscope. As its primary physics goal, IAXO will look for axions or axion-like particles (ALPs) originating in the Sun via the Primakoff conversion of the solar plasma photons. In terms of signal-to-noise ratio, IAXO will be about 4-5 orders of magnitude more sensitive than CAST, currently the most powerful axion helioscope, reaching sensitivity to axion-photon couplings down to a few $\times 10^{-12}$~GeV$^{-1}$ and thus probing a large fraction of the currently unexplored axion and ALP parameter space. IAXO will also be sensitive to solar axions
produced by mechanisms mediated by the axion-electron coupling $g_{ae}$ with sensitivity $-$for the first time$-$ to values of $g_{ae}$ not previously excluded by astrophysics. With several other possible physics cases, IAXO has the potential to serve as a multi-purpose facility for generic axion and ALP research in the next decade.
In this paper we present the conceptual design of IAXO, which follows the layout of an enhanced axion helioscope, based on a purpose-built 20m-long 8-coils toroidal superconducting magnet. All the eight 60cm-diameter magnet bores are equipped with focusing x-ray optics, able to focus the signal photons into $\sim 0.2$~cm$^2$ spots that are imaged by ultra-low-background Micromegas x-ray detectors. The magnet is built into a structure with elevation and azimuth drives that will allow for solar tracking for $\sim$12 h each day.

}


\begin{document}

\section{Introduction}\label{Introduction}

Axions appear~\cite{Weinberg:1977ma,Wilczek:1977pj} in very well motivated extensions of the Standard Model (SM) including the Peccei-Quinn mechanism~\cite{Peccei:1977hh,Peccei:1977ur} proposed to solve the long-standing strong-CP problem~\cite{Cheng:1987gp}. Together with the weakly interacting massive particles (WIMPs) of supersymmetric theories, axions are also favored candidates to solve the Dark Matter (DM) problem~\cite{Bae:2008ue,Visinelli:2011wa}. Their appeal comes from the fact that, like WIMPs, they are not an \textit{ad hoc} solution to the DM problem. Mixed WIMP-axion DM is one possibility favored in some theories~\cite{Baer:2011uz,Bae:2013pxa}. More generic axion-like particles (ALPs) appear in diverse extensions of the SM (e.g., string theory)~\cite{Arvanitaki:2009fg,Cicoli:2012sz,Ringwald:2012cu}. ALPs could also be the DM~\cite{Arias:2012az} and are repeatedly invoked to explain some astrophysical observations.


The diverse experimental approaches to search for axions can be classified in three main categories, complementary on many levels~\cite{ANDP:ANDP201300727,Carosi:2013rla}: \emph{haloscopes}~\cite{Sikivie:1983ip} look for the relic axions potentially composing our dark matter galactic halo, \emph{helioscopes}~\cite{Sikivie:1983ip} look for axions emitted at the core of the sun, and \emph{light-shining-through-wall} (LSW) experiments~\cite{Redondo:2010dp} look for axion-related phenomena generated entirely in the laboratory. All three strategies invoke the generic axion-photon interaction, a necessary property of axions, and thus rely on the use of powerful magnetic fields to trigger the conversion of the axions into photons that can be subsequently detected. Among these approaches the axion helioscope stands out as the most mature, technologically feasible and capable of being scaled in size.

The most relevant channel of axion production in the solar core is the Primakoff conversion of plasma photons into axions in the Coulomb field of charged particles via the generic $a\gamma\gamma$ vertex. The Primakoff solar axion flux peaks at 4.2 keV and exponentially decreases for higher energies~\cite{Andriamonje:2007ew}. This spectral shape is a robust prediction depending only on well known solar physics, while the only unknown axion parameter is the axion-photon coupling constant $\gagamma$ and enters the flux as an overall multiplicative factor $\propto\gagamma^2$. For the particular case of non-hadronic axions having tree-level interactions with electrons, other productions channels (e.g., brehmstrahlung, compton or axion recombination) should be taken into account, as their contribution can be greater than that of the Primakoff mechanism~\cite{Barth:2013sma,Redondo:2013wwa}.


The basic layout of an axion helioscope thus requires a powerful magnet
coupled to one or more x-ray detectors. When the magnet is aligned
with the Sun, an excess of x-rays at the exit of the magnet is
expected, over the background measured at non-alignment periods. This detection concept was first experimentally realized
at Brookhaven National Laboratory (BNL) in 1992. A stationary dipole magnet with a field of $B = 2.2$~T and a length of $L = 1.8$~m was oriented towards the setting Sun~\cite{Lazarus:1992ry}. The experiment derived an upper limit on $\gagamma$
$(99\%$ CL) $< 3.6\times10^{-9}$ GeV$^{-1}$ for $m_a < 0.03$ eV. At the University of Tokyo, a second-generation experiment was built: the Tokyo axion heliscope (also nicknamed Sumico). Not only did this experiment implement a dynamic tracking of the Sun but it also used a more powerful magnet ($B =$ 4 T, $L =$ 2.3 m) than the BNL predecessor. The bore, located between the two coils of the magnet, was evacuated and higher-performance detectors were installed~\cite{Inoue:2002qy,Moriyama:1998kd,Inoue:2008zp}. This new setup resulted in an improved upper limit in the mass range up to 0.03 eV of $\gagamma (95\%~{\rm CL} ) < 6.0\times 10^{-10}$ GeV$^{-1}$. Later experimental improvements included the additional use of a buffer gas to enhance sensitivity to higher-mass axions.

A third-generation experiment, the CERN Axion Solar Telescope (CAST), began data collection in 2003. The experiment uses a Large Hadron Collider (LHC) dipole prototype magnet with a magnetic field of up to 9 T over a length of 9.3 m~\cite{Zioutas:1998cc}. Like Sumico, CAST is able to follow the Sun for several hours per day using a sophisticated elevation and azimuth drive. This CERN experiment is the first helioscope to employ x-ray focusing optics for one of its four detector lines~\cite{Kuster:2007ue}, as well as low background techniques from detectors in underground laboratories~\cite{Abbon:2007ug}. During its observational program from 2003 to 2011, CAST operated first with the magnet bores in vacuum (2003--2004) to probe masses $m_a < 0.02$~eV. No significant signal above background was observed. Thus, an upper limit on the axion-to-photon coupling of $\gagamma~(95\%~$CL$) < 8.8\times 10^{-11}$ GeV$^{-1}$ was obtained~\cite{Zioutas:2004hi,Andriamonje:2007ew}. The experiment was then upgraded to be operated with $^4$He (2005--2006) and
$^3$He gas (2008--2011) to obtain continuous, high sensitivity up to an axion mass of $m_a = 1.17$ eV. Data released up to now provide an average limit of $\gagamma~(95\%~$CL$) \lesssim 2.3\times10^{-10}$ GeV$^{-1}$, for the higher mass range of 0.02 eV $< m_a <$ 0.64 eV~\cite{Arik:2008mq,Aune:2011rx} and of about $\gagamma~(95\%~$CL$) \lesssim 3.3\times10^{-10}$ GeV$^{-1}$ for 0.64 eV $< m_a <$ 1.17 eV~\cite{Arik:2013nya}, with the exact value depending on the pressure setting.

So far each subsequent generation of axion helioscopes has resulted in an improvement in sensitivity to the axion-photon coupling constant of about a factor 6 in $\gagamma$ over its predecessors. CAST has been the first axion helioscope to surpass the stringent limits from astrophysics $\gagamma \lesssim 10^{-10}$ GeV$^{-1}$ over a large mass range and to probe previously unexplored ALP parameter space. In particular, in the region of higher axion masses ($m_a \gtrsim 0.1$ eV), CAST has entered the band of QCD axion models for the first time and excluded KSVZ axions of specific mass values. We have shown~\cite{Irastorza:2011gs} that a further substantial step beyond the current state-of-the-art represented by CAST is possible with a new fourth-generation axion helioscope. This concept has been recently materialized in the International Axion Observatory (IAXO), recently proposed to CERN~\cite{Irastorza:1567109}, whose conceptual design is the subject of this paper.

As its primary physics goal, IAXO will look for axions or ALPs originating in the Sun via the Primakoff conversion of the solar plasma photons. In terms of signal-to-background ratio, IAXO will be about 4$-$5 orders of magnitude more sensitive than CAST, which translates into a factor of $\sim$20 in terms of the axion-photon coupling constant $\gagamma$. That is, this instrument will reach the few $\times$10$^{-12}~{\rm GeV}^{-1}$ regime for a wide range of axion masses up to about 0.25~eV.
IAXO has potential for the discovery of axions and other ALPs,
since it will deeply enter into completely unexplored parameter space.
At the very least it will firmly exclude a huge region of this space.
Needless to say, the discovery of such particles and the consequent evidence for
physics at very high energy scales would be a groundbreaking result for particle physics.

\begin{figure}[t] \centering
\includegraphics[width=\textwidth]{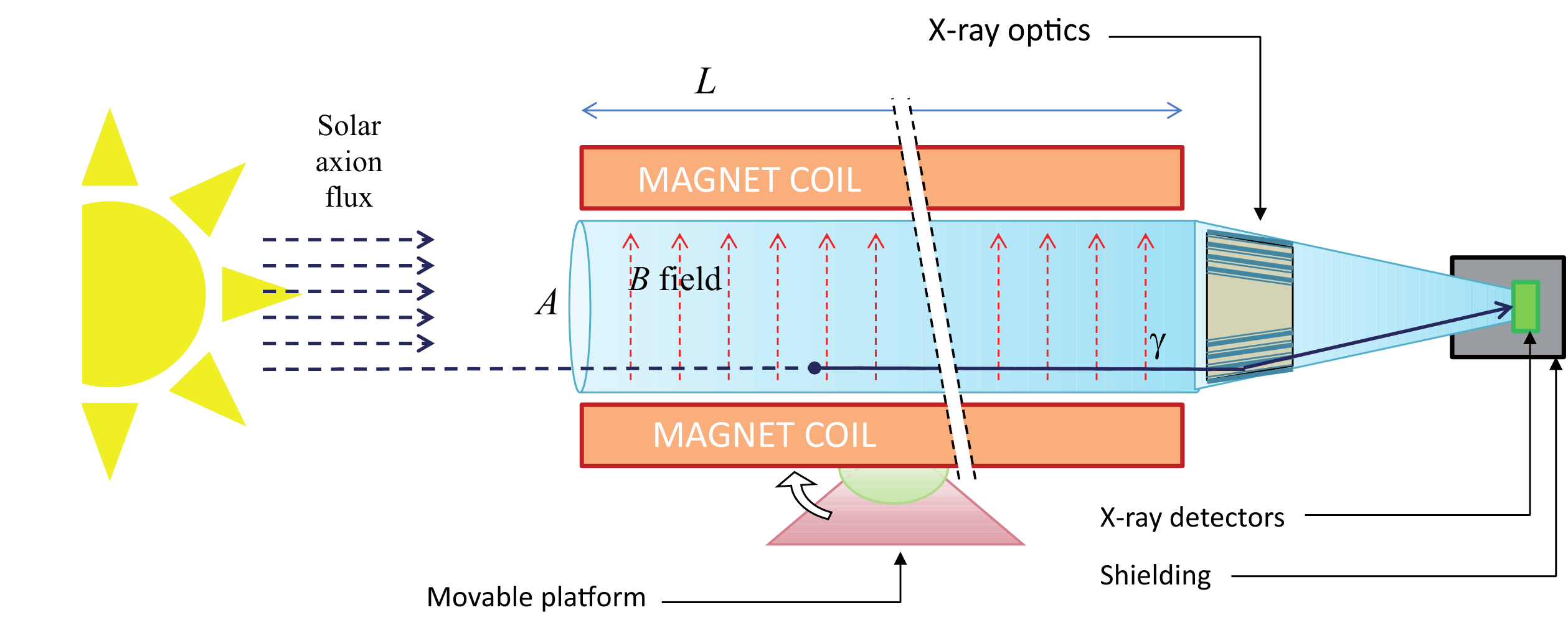}\hspace{2pc}%
\caption{\label{fig:NGAH_sketch} Conceptual arrangement of an enhanced axion helioscope with x-ray focalization. Solar axions are converted into photons by the transverse magnetic field inside the bore of a powerful magnet. The resulting quasi-parallel beam of photons of cross sectional area $A$ is concentrated by an appropriate x-ray optics into a small spot area $a$ in a low background detector. The envisaged implementation in IAXO (see figure~\protect\ref{fig:IAXO_sketch}, includes eight such magnet bores, with their respective optics and detectors. }
\end{figure}

In order to achieve the stated sensitivity, IAXO follows the conceptual layout of an enhanced axion helioscope~\cite{Irastorza:2011gs}, sketched in figure~\ref{fig:NGAH_sketch}, in which all the magnet aperture is coupled to focusing optics. It relies on the construction of a large superconducting 8-coil toroidal magnet optimized for axion research. Each of the eight 60 cm diameter magnet bores is equipped with x-ray optics focusing the signal photons into $\sim$0.2 cm$^2$ spots that are imaged by ultra-low background Micromegas x-ray detectors. The magnet will be built into a structure with elevation and azimuth drives that will allow solar tracking for $\sim$12 hours each day. All the enabling technologies for IAXO exist, there is no need for development. IAXO will also benefit from the invaluable expertise and knowledge gained from the successful operation of CAST for more than a decade.

We refer to \cite{Irastorza:2011gs} for a description of the first motivation and the figure-of-merit study that supports the IAXO concept. A detailed study of the physics potential of IAXO will be included in a paper currently under preparation, although it can also be found in the Letter of Intent recently submitted to CERN~\cite{Irastorza:1567109}. In the following sections we describe the different parts of IAXO, focusing on the enabling technologies of the experiment. The toroidal superconducting magnet is described in section~\ref{sec:magnet}. The IAXO x-ray focusing optics are described in section~\ref{sec:optics}. The Micromegas low-background detectors are described in section~\ref{sec:detectors}. In section~\ref{sec:additional} the main features of the experiment's tracking platform, as well as potential additional equipment are briefly described. Finally, we conclude with section \ref{sec:conclusions}.

\vspace{2mm}

\section{The IAXO superconducting magnet}
\label{sec:magnet}
The outcome of the figure of merit (FOM) analysis~\cite{Irastorza:2011gs} indicates the importance and need for a new magnet to achieve a significant step forward in the sensitivity to the axion-photon coupling. The design of the new magnet is performed with the magnet's FOM (MFOM) in mind already from the initial design stages. Since practically and cost-wise the currently available detector (i.e. large scale) magnet technology is limited to using NbTi superconductor technology which allows peak magnetic field of up to 5-6~T, the magnet's aperture is the only MFOM parameter that can be considerably enlarged. Consequently, the design of the magnet has started with the focus on this parameter. The preliminary optimization study
has shown that the toroidal geometry is preferred for an axion helioscope \cite{Irastorza:2011gs}. Inspired by the ATLAS barrel and end-cap toroids, a large superconducting toroidal magnet is currently being designed to fulfill the requirements of IAXO.
The new toroid will be built up from eight, 1m-wide and 21m-long, racetrack coils. The innovative magnet system is sized 5.2~m in diameter and 25~m in length. It is designed to realize a peak magnetic field of 5.4~T with a stored energy of 500~MJ at the operational current of 12.3~kA.


\begin{figure}[!b]
\begin{center}
    \includegraphics[width=\textwidth] {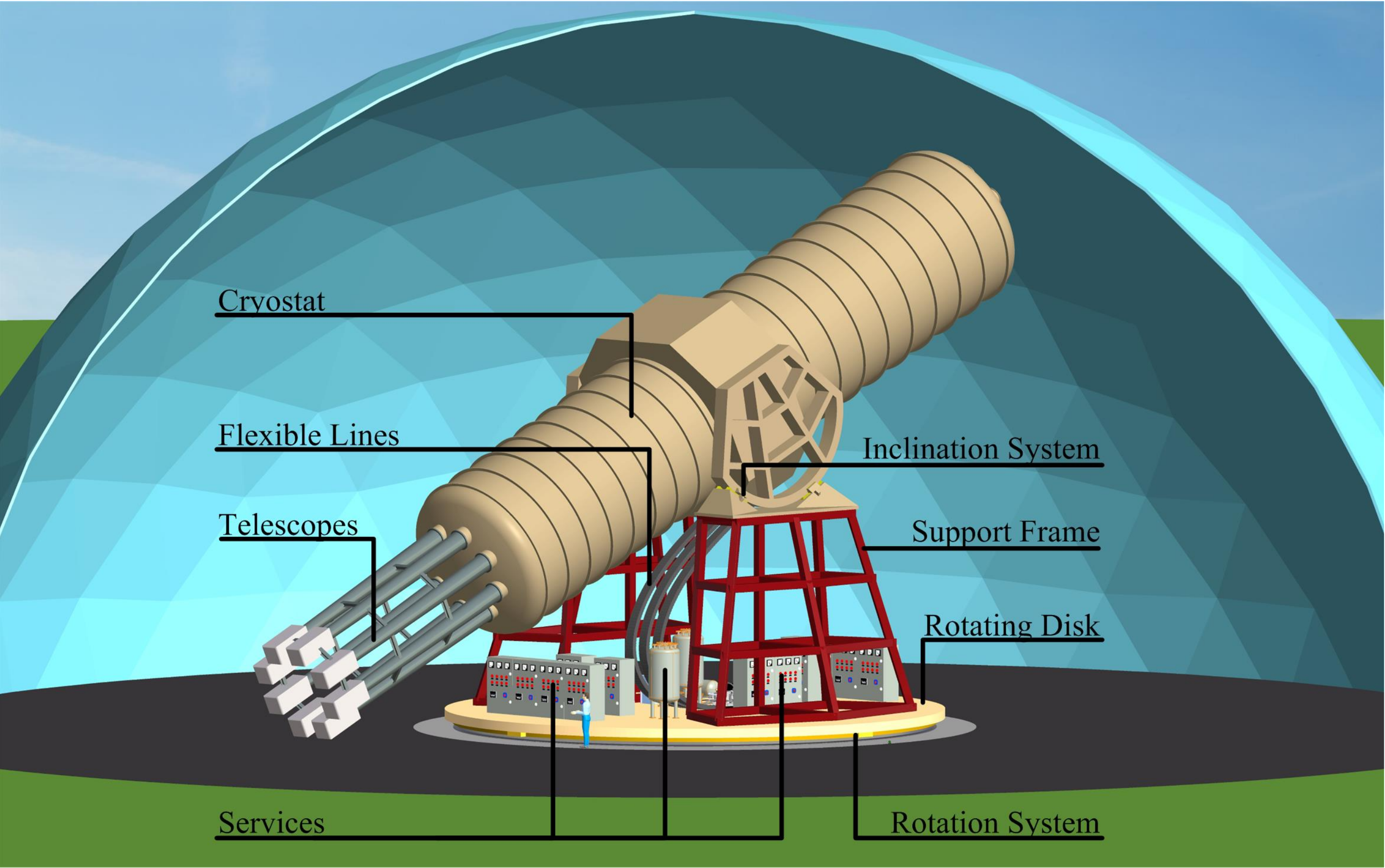}
    \caption{Schematic view of IAXO. Shown are the cryostat, eight x-ray optics and detectors, the flexible lines guiding services into the magnet, cryogenics and powering services units, inclination system and the rotating disk for horizontal movement. The dimensions of the system can be appreciated by a comparison to the human figure positioned by the rotating table.}
    \label{fig:IAXO_sketch}
\end{center}
\end{figure}

\renewcommand{\arraystretch}{1}
\begin{table}[!t]
\begin{center}
\begin{tabular*}{0.8\textwidth}{@{\extracolsep{\fill}} p{9.5 cm}  cc} \hline
\textit{Property} & \textit{Value} & \textit{Unit}\\
\hline
\textbf{Cryostat dimensions:} \hfill Overall length & 25 & m  \\
\hfill Outer diameter & 5.2 & m \\
\hfill Cryostat volume & $\sim$ 530 & m$^3$\\
\textbf{Toroid size:} \hfill Inner radius, $R_{in}$  & 1.05 & m  \\
\hfill Outer radius, $R_{out}$ & 2.05 & m \\
\hfill Inner axial length & 21.0 & m\\
\hfill Outer axial length & 21.8 & m\\
\textbf{Mass:} \hfill Conductor & 65 & tons\\
\hfill Cold Mass  & 130 & tons\\
\hfill Cryostat  & 35 & tons\\
\hfill Total assembly   & $\sim$ 250 & tons\\

\textbf{Coils:}  \hfill Number of racetrack coils& 8 & -\\
\hfill Inner radius of bare coil, relative to racetrack center & 500 & mm\\
\hfill Outer radius of bare coil, relative to racetrack center & 884 & mm\\
\hfill Inner winding radius in corner & 500 & mm\\
\textit{Winding dimensions:}
\hfill Winding pack width & 384 & mm \\
\hfill Winding pack height & 144 & mm \\
\hfill Length of inner turn & 43.1 & m\\
\hfill Length of outer turn & 45.5 & m\\
\hfill Turns/coil & 180 & -\\

\textit{Nominal Values:} \hfill Nominal current, $I_{op}$ & 12.3 & kA\\
\hfill Stored energy, $E$  & 500 & MJ\\
\hfill Inductance & 6.9 & H\\
\hfill Peak magnetic field, $B_p$ & 5.4 & T\\
\hfill Average field in the bores & 2.5 & T \\

\textbf{Conductor:} \hfill Conductor unit length per double-pancake & 4.0 & km\\
\hfill Conductor length per coil & 8.0 & km\\
\hfill Total conductor length (including reserve) & 68 & km\\
\hfill Cross-sectional area & 35 $\times$ 8 & mm$^2$\\
\hfill Number of strands & 40 & -\\
\hfill Strand diameter & 1.3 & mm\\
\hfill Critical current @ 5 T, $I_c$ & 58 & kA\\
\hfill Operating temperature, $T_{op}$ & 4.5 & K\\
\hfill Operational margin & 40\% & -\\
\hfill Temperature margin @ 5.4 T  & 1.9 & K\\
\textbf{Heat Load:} \hfill  at 4.5 K & $\sim$150 & W\\
\hfill at 60-80 K  & $\sim$1.6 & kW\\
\hline \end{tabular*}
\end{center}
\label{table1} \caption{Main design parameters of the IAXO toroidal magnet.}
\end{table}

\subsection{Figure of merit and lay-out optimization}

The general guideline to define the lay-out of the new toroidal magnet has been to optimize the MFOM $f_M = L^2B^2A$, as defined in~\cite{Irastorza:2011gs}, where $L$ is the magnet length, $B$ the effective magnetic field and $A$ the aperture covered by the x-ray optics. Currently, the MFOM of the CAST magnet is 21~T$^2$m$^4$. As discussed in \cite{Irastorza:2011gs}, an MFOM of 300 relative to CAST is necessary for IAXO to aim at sensitivities to $\gagamma$ of at least one order of magnitude beyond the current CAST bounds. Accordingly, we have adopted the latter value as the primary design criterion for the definition of the toroidal magnet system, together with other practical constraints such as the maximum realistic size and number of the x-ray optics (section~\ref{sec:optics}) and the fact that the design should rely on known and well proven engineering solutions and manufacturing techniques.

\begin{figure}[!t]
\begin{center}
	\includegraphics[scale=0.12]{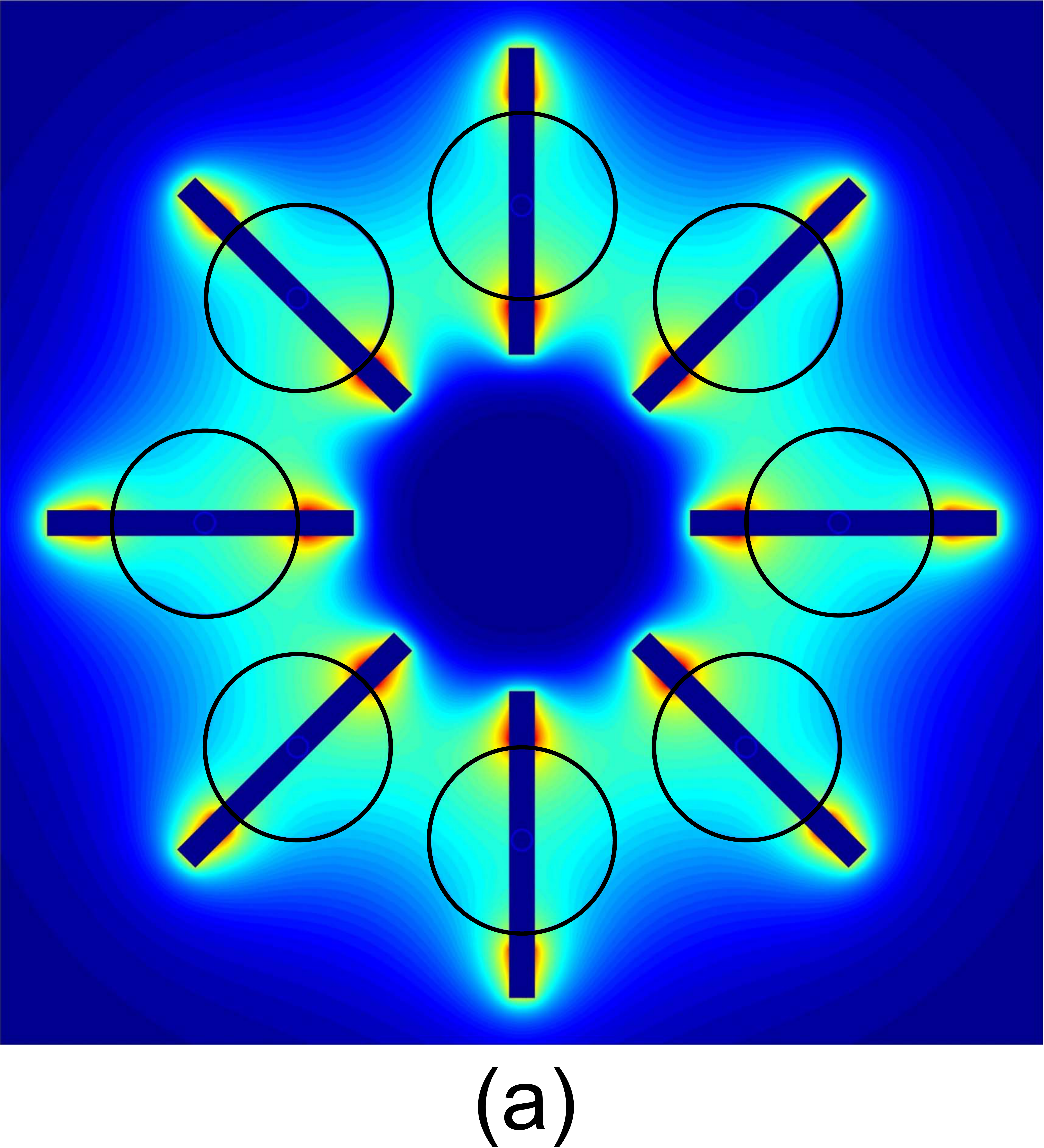}
	\includegraphics[scale=0.12]{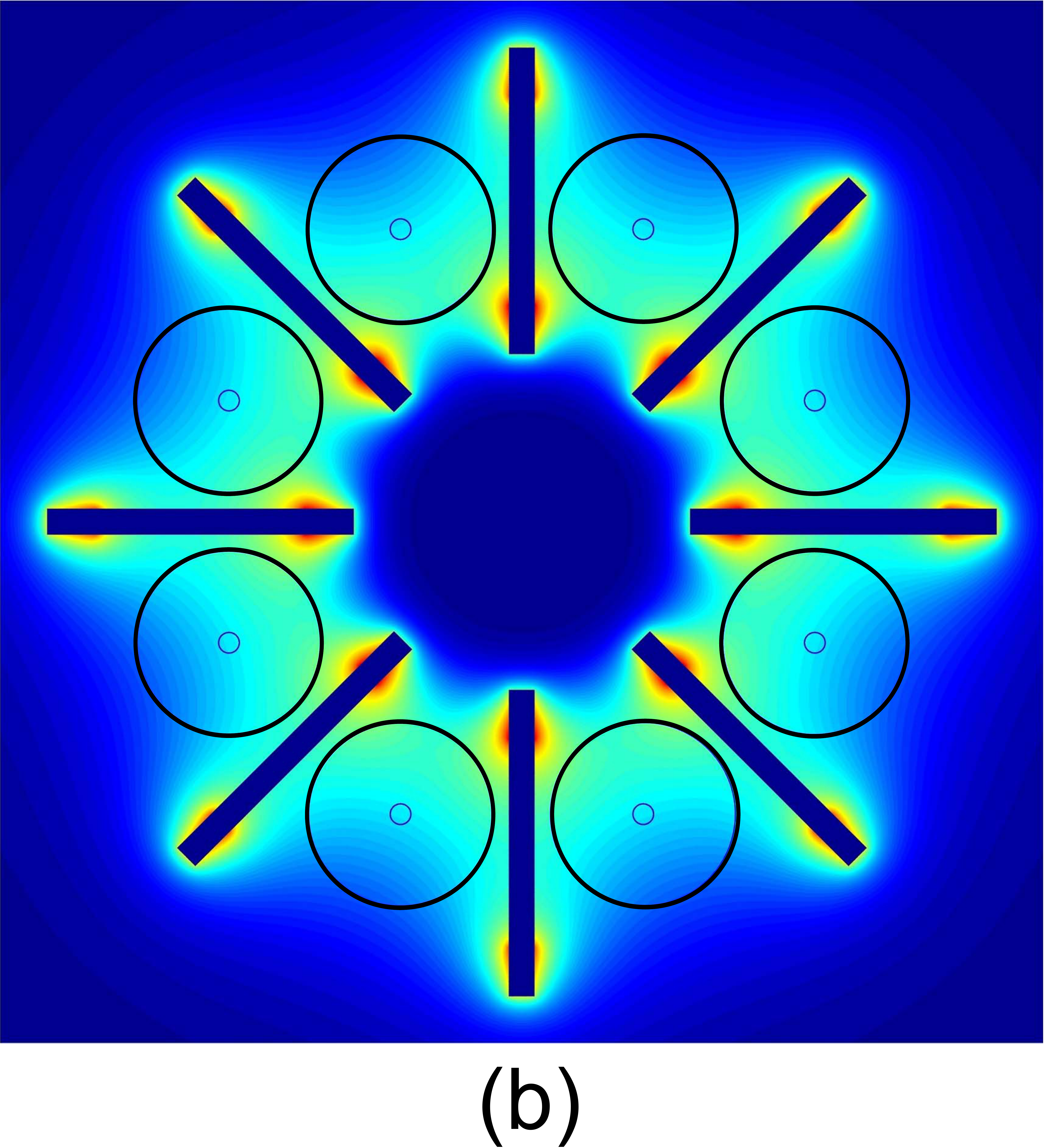}
	\caption{Illustration of the two principle angular alignment options considered for the optics with respect to the coils. The rectangles represent the toroid's coil and the circles represent the optics' bores. (a) "Field dominated" alignment: optics behind the coils. (b) "Area dominated" alignment: optics in between the coils.}
	\label{alignment}
\end{center}
\end{figure}

To determine the MFOM, the magnet straight section length $L$ is set to 20~m and the integration $\int B^2(x,y)dxdy$ is performed over the \textit{open} area covered by the x-ray optics. Hence, to perform the integration, the optics' positioning must be determined. Upon placing the optics as close as possible to the inner radius of the toroid $R_{in}$, the optimized angular alignment of the optics is determined by the result of the integration. Two principal options for the angular alignment are considered: one is to align each of the optics between each pair of racetrack coils, whereas the other is to place the optics behind the racetrack coils. Fig. \ref{alignment} provides a general illustration of the two alignment options for an 8-coils toroid. In practice, the two options represent two different approaches: the first, referred to as the "area dominated" option, takes advantage of the entire large aperture of each of the optics and the second "field dominated" option assumes that placing the coils behind the optics, and by that including areas with higher magnetic field in the integration, will increase $f_M$.

The magnetic field is determined by the geometrical and electromagnetic parameters of the magnet. For each lay-out, the magnetic field is calculated using a 3D finite element analysis (FEA) model and the integration is performed on the mid-plane. Once the position of the optics is fixed, the integration over a disc with radius $R_{det}$ centered at $(R_{cen}, \theta_{cen})$ can be performed. The model features an arc at the bent sections of each racetrack with a radius $R_{arc} = (R_{out} - R_{in})/2$, where $R_{out}$ and $R_{in}$ are the outer and inner radii of the racetrack coil windings, respectively. The model also assumes the use of an Al stabilized Rutherford NbTi cable in the coil windings. The winding dimensions are determined from the conductor dimensions assuming a few winding configurations.

The optimization study shows that IAXO's MFOM is affected considerably by the fraction of the aperture of the optics exposed to x-rays, thus favoring the area dominated alignment. Even when considering the field dominated alignment, it is preferable to use thinner coils, thus increasing the open aperture in front of the optics. Moreover, the area dominated option yields a $15 \%$ larger MFOM, compared to the field dominated option.

The magnet system design, presented in Fig. \ref{fig:IAXO_sketch}, follows the result of the geometrical optimization study. The design meets all the experimental requirements of the magnet. It is relying on known and mostly well-proven engineering solutions, many of which were used in and developed for the ATLAS toroids engineered by CERN, INFN Milano and CEA Saclay. This ensures that the magnet is technically feasible to manufacture. The main properties of the toroid are listed in Table 1. The design essentially features a separation of the magnet system from the detection systems, which considerably simplifies the overall system integration. This also allows for eight open bores, which are centered and aligned in between the racetrack coils, in accordance with the geometrical study. The inclusion of the eight open bores will simplify the fluent use of experimental instrumentation\footnote{An exception to this may be the use of microwave cavities (see section~\ref{sec:additional_mw}), which could profit from a cryogenic environment to achieve low levels of noise.}.

The toroidal magnet comprises eight coils and their casing, an inner cylindrical support for the magnetic forces, keystone elements to support gravitational and magnetic loads, a thermal shield, a vacuum vessel and a movement system (see Figs. \ref{fig:IAXO_sketch} to \ref{cs} and Table 1). Its mass is $\sim$250~tons. At the operational current of 12.3~kA the stored energy is $\sim$500~MJ. The design criteria for the structural design study are defined as: a maximum deflection of 5 mm, a general stress limit of 50~MPa and a buckling factor of 5. The magnetic and structural designs are done using the ANSYS$^{\textregistered}$ 14.5 Workbench environment. The Maxwell 16.0 code is used to calculate 3D magnetic fields and Lorentz forces. The magnetic force load is linked to the static-structural branch to calculate stress and deformation. The eight bores are facing eight X-ray optics with a diameter of 600~mm and a focal length of $\sim$6~m. The diameter of the bores matches the diameter of the optics.

It is worth mentioning that numerous other magnet designs (e.g. accelerator magnets, solenoids and dipole structures) were considered during the optimization study as well \cite{Irastorza:2011gs}. Also, less conventional toroidal designs were examined. For example, ideas for racetrack windings with bent ends were suggested to reduce the area loss when implementing the field dominated alignment. For the same reason toroids with slightly tilted coils were discussed. In general, these designs pose significant technical complications while offering a low potential to significantly enlarge the MFOM and hence deviate from the philosophy behind the magnet concept. In addition, toroidal geometries with more coils and bores were studied essentially to enhance the detection area. The MFOM scales linearly with the number of coils (when keeping the optics' cross-section constant), which points out that the choice for an eight toroid is cost driven in essence.

\subsection{Conductor}

\begin{figure}
\begin{center}
    \includegraphics[height=7cm] {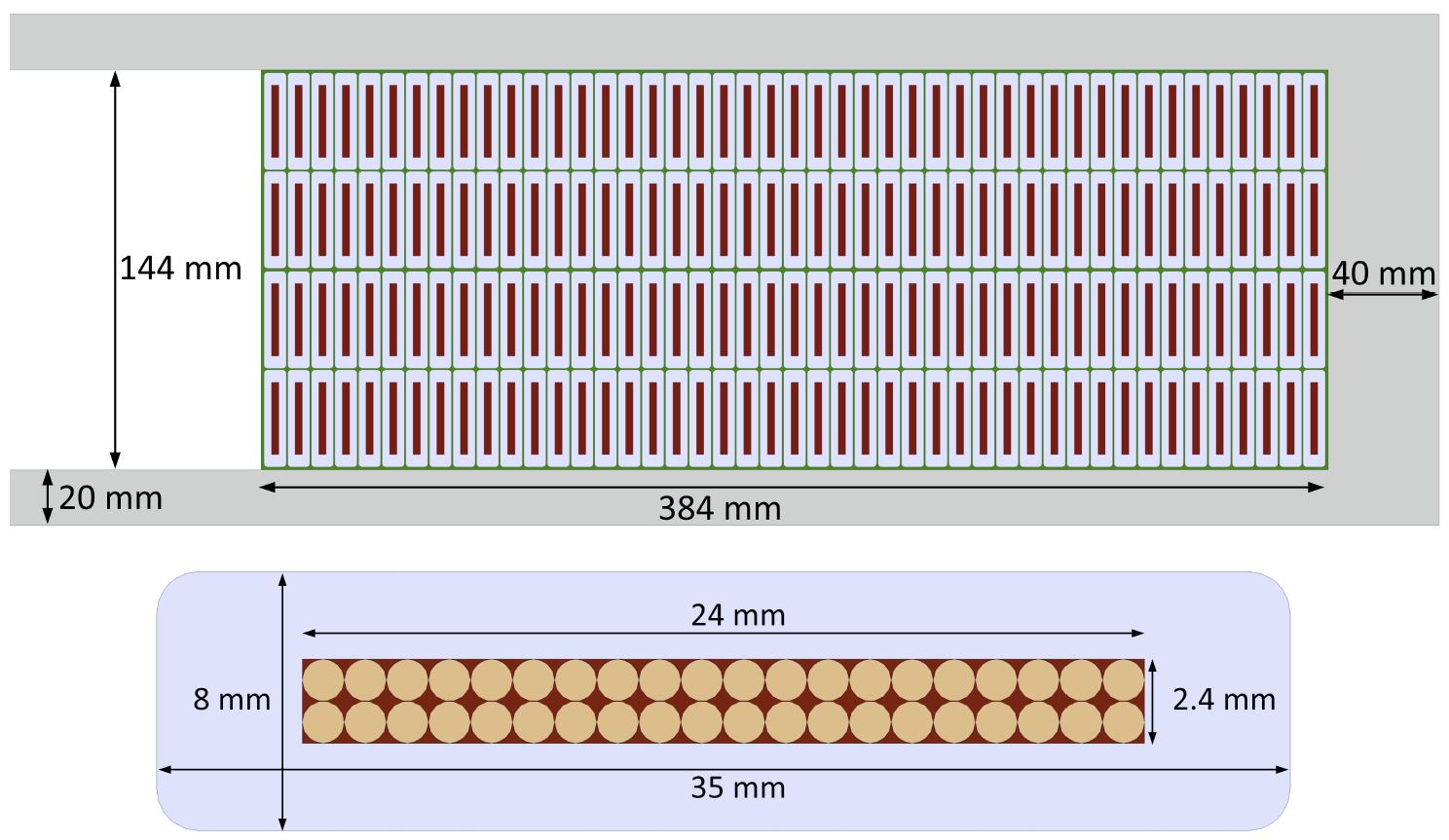}
    \caption{Cross section of the two double pancake winding packs, the coil casing (top) and the conductor with a 40 strands NbTi/Cu Rutherford cable embedded in a dilute Al-0.1wt\%Ni doped stabilizer (bottom).}
    \label{coil}
\end{center}
\end{figure}

The conductor is shown in Fig.~\ref{coil}. The Rutherford type NbTi/Cu cable, composed of 40 strands of 1.3~mm diameter and a Cu/NbTi ratio of 1.1, is co-extruded with a Al-0.1wt\%Ni stabilizer with high residual-resistivity ratio, following the techniques used in the ATLAS and CMS detector magnets~\cite{tenKate:1158687,tenKate:1169275,Herve:1169279}. The use of a Rutherford cable as the superconducting element provides a high current density while maintaining high performance redundancy in the large number of strands. The Al stabilizer serves both quench protection and stability for the superconductor. The conductor has a critical current of $I_c(5~\mbox{T},~ 4.5~\mbox{K}) =$ 58~kA.

\subsubsection{Peak magnetic field and forces}

The peak magnetic field in the windings, which determines the operation point of the conductor and the temperature margin, is calculated at 5.4~T for a current of 12.3~kA per turn. In order to minimize the forces acting on the bent sections, the racetrack coils are bent to a symmetric arc shape, with $R_{arc} = 0.5$~m. The net force acting on each racetrack coil is 19~MN, directed radially inwards.

\subsubsection{Stability analysis}

The IAXO magnet requires maintaining the highest possible magnetic field in order to maximize the MFOM. However, suitable operational current and temperature margins are mandatory to ensure its proper and safe operation. For a two double pancake configuration with 180 turns and engineering current density $J_{eng} = 40$~A/mm$^2$, the peak magnetic field is $B_p = 5.4$~T. Following this, the critical magnetic field corresponding to the magnet load line is 8.8~T at 65~A/mm$^2$. Hence, IAXO's magnet is working at about 60\% on the load line, setting the operational current margin to 40\%.

The temperature margin calculation is based on an operational temperature of $T_{op} = 4.5$~K and a peak magnetic field of $B_p = 5.4$~T. A coil with two double pancakes and 45 turns per pancake satisfies this requirement with a temperature margin of 1.9~K, while yielding an MFOM of 300, relative to CAST, thus satisfying the principal design criterion.

\subsection{Electrical circuit and quench protection}

The adiabatic temperature rise in the case of a uniformly distributed quench is $\sim$ 100~K. The toroid's quench protection is based on an active system and an internal dump of the stored energy. The principle of the quench protection system is to rely on simple, robust and straightorward detection circuits and electronics and have sufficient redundancy in order to reduce failure probability.

\begin{figure}[!t]
\centering
    \includegraphics[width=0.9\textwidth] {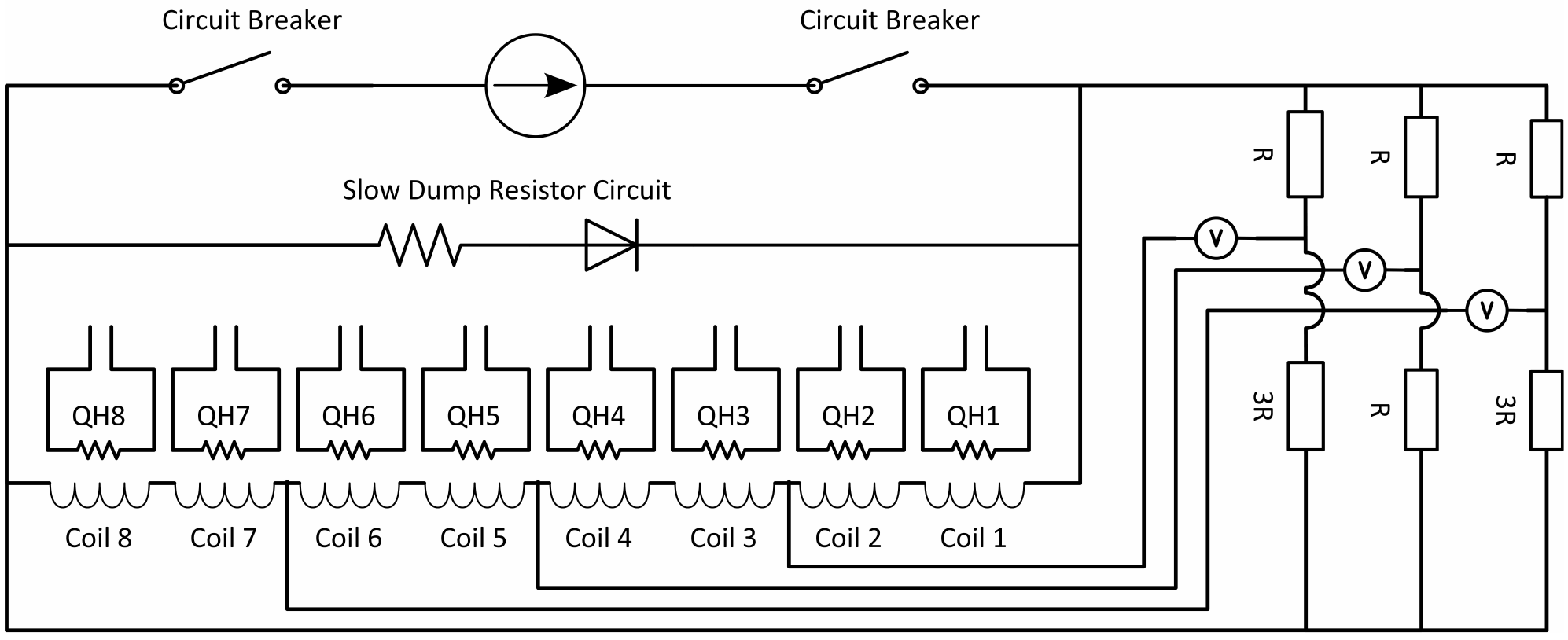}
    \caption{Schematic diagram of the electrical circuit and quench protection scheme. Shown are the power convertor, the eight coils, quench heaters (QH 1-8), the slow dump circuit and the quench detection circuit.}
    \label{el}
\end{figure}

The electrical circuit of the IAXO toroid is shown in Fig.~\ref{el}. The magnet power convertor of 12.5~kA is connected at its DC outputs to two breakers, which open both electrical lines to the magnet. The high-$T_c$ current leads are installed within their own cryostat, that is integrated on the rotating gantry of the magnet. The current leads feed the eight coils, which are connected in series, by means of flexible superconducting cables. The flexibility is required to compensate for the changing inclination of the racetrack coils. Each coil is equipped with multiple quench heaters, connected in parallel. Across the warm terminals of the current leads, a slow-dump-resistor with low resistance is connected in series to a diode. The quench detection circuit relies on the detection of normal voltage growth across the toroid following a quench. The voltage sensitivity level of the detectors is 0.3~V, which implies a typical detection delay time of $\sim$ 1~s. The protection circuit is equipped with a timer delay so that false signals do not activate the protection system and lead to a bogus fast discharge.

When a quench is detected and verified, the two breakers open to quickly separate the magnet from the power convertor and a quench is initiated in all coils simultaneously by activating all the quench heaters. This ensures a fast and uniform quench propagation and thus a homogenous cold mass temperature after a quench. Simultaneously, the current is discharged through the dump-resistor. This discharge mode, the so-called fast dump mode, is characterized by an internal dump of the magnet's stored energy, because the magnetic energy is dissipating into heat in the magnet windings. Upon grounding the magnet, the fast discharge scheme ensures that the discharge voltage excitation is kept low enough and that the stored energy is uniformly dissipated in the windings. The internal energy dump depends on the absolute reliability of the quench heaters system. To reduce failure probability to an acceptable level, the quench protection system features a six-fold redundant quench detection circuit with bridges and a two-fold redundant quench heater system with multiple heaters along each of the eight racetracks.

The DC power convertor will operate in voltage control mode when ramping up the toroid and in current control mode during steady operation. The field stability requirement for an axion helioscope is of minor importance. A time variability as large as 0.1\% will not affect the axion-photon conversion probability, and hence the experiment's sensitivity.

Under normal operation, the toroid will be discharged through the diode-resistor circuit in a passive run down mode (slow discharge mode). Slow discharge is also the safety dump mode activated in the case of a minor fault.

Each of the dump-resistors is connected in series to a diode unit to avoid current driven through the dump resistor circuit during normal operation of the magnet. The dump resistors circuit is air cooled by convection and have the capacity of absorbing the total stored magnetic~energy~of~the~toroid.

The longitudinal normal zone propagation velocity is $\sim$6.5~m/s. The velocity was calculated by using COMSOL 4.3b coupled multiphysics modules~in~a~2D adiabatic model. Hence, the normal zone will propagate around an entire coil (43~m) in 3.3~s.

\subsection{Cold-mass}

The cold mass operating temperature is 4.5~K and its mass is approximately 130~tons. The cold mass consist of eight coils, with two double pancakes per coil, which form the toroid geometry, and a central cylinder designed to support the magnetic force load. The coils are embedded in Al5083 alloy casings, which are attached to the support cylinder at their inner edge. The casings are designed to minimize coil deflection due to the magnetic forces.

\begin{figure}[b!]
\begin{center}
    \includegraphics[height=10cm] {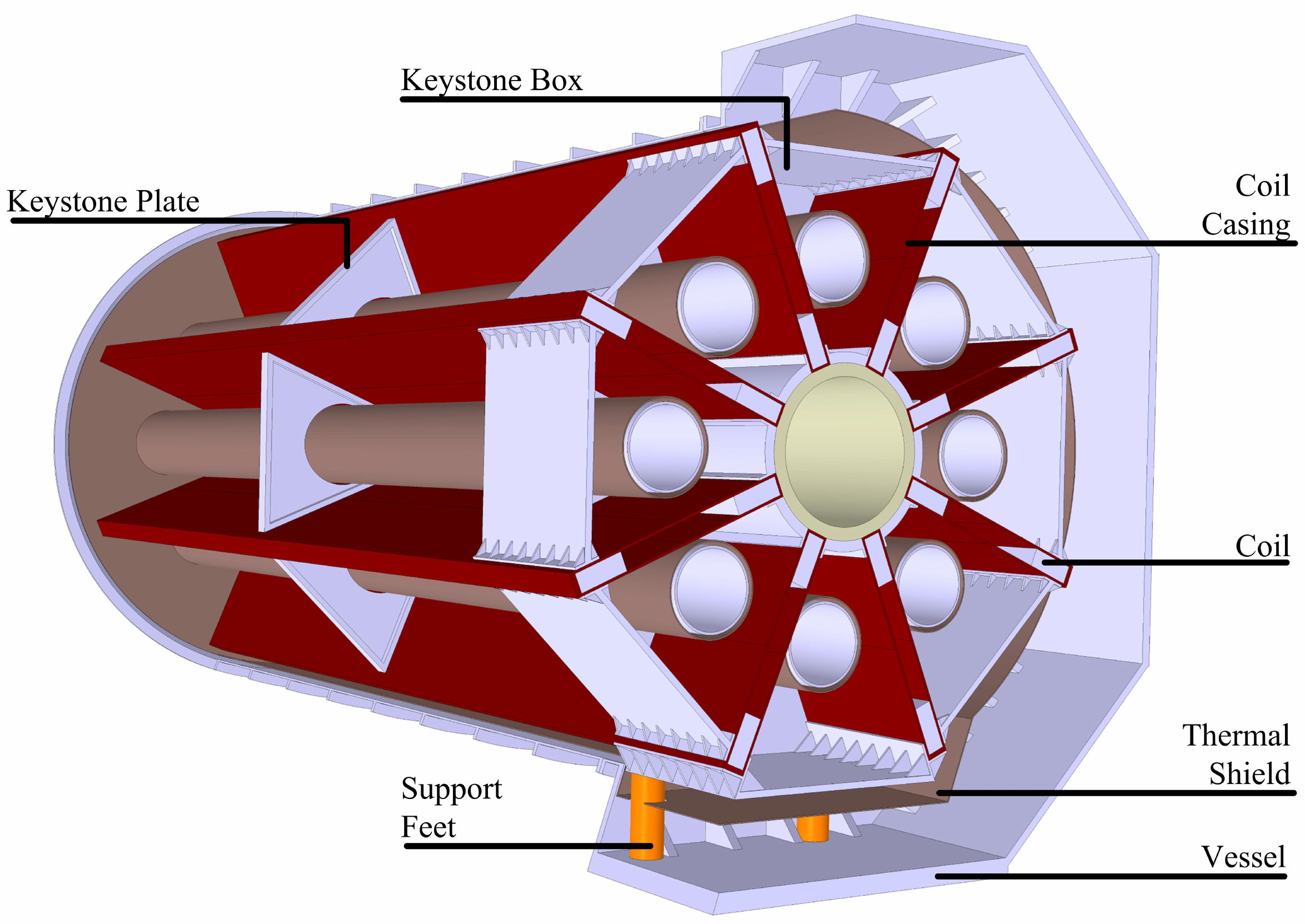}
    \caption{Mid-plane cut of the cryostat with an exposed cold mass, showing the cold mass and its supports, surrounded by a thermal shield, and the vacuum vessel. The open bores will simplify the use of experimental instrumentation.}
    \label{cs}
\end{center}
\end{figure}

To increase the stiffness of the cold mass structure and maintain the toroidal shape under gravitational and magnetic loads, and to support the warm bores, eight Al5083 keystone boxes and 16 keystone plates are connected in between each pair of coils, as shown in Fig. \ref{cs}. The keystone boxes are attached to the support cylinder at the center of mass of the \emph{whole} system (i.e. including the optics and detectors) and the keystone plates are attached at half-length between the keystone boxes and the coils ends.

A coil, shown in Fig. \ref{coil}, comprises two double pancake windings separated by a 1 mm layer of insulation. The coils are impregnated for proper bonding and pre-stressed within their individual casing to minimize shear stress and prevent cracks and gaps appearing due to thermal shrinkage on cool-down and magnetic forces.

\subsection{Cryostat and its movement system}

The design of the cryostat is based on a rigid central part, placed at the center of mass of the whole system and serves as a fixed support point of the cold mass, with two large cylinders and two end plates enclosing it to seal the vacuum vessel. In addition, eight cylindrical open bores are placed in between the end plates. The vessel is optimized to sustain the atmospheric pressure difference and the gravitational load, while being as light and thin as possible. The Al5083 rigid central piece is 70~mm thick with a thicker 150~mm bottom plate to support the cold mass. Using two end flanges at the vessel's rims, as well as periodic reinforcement ribs at 1.35~m intervals along both cylinders, the structural requirements are met for a 20~mm thick Al5083 vessel with two 30~mm thick torispherical Kl\"opper shaped end plates. The 10~mm wall thickness of the eight cylindrical bores is minimized in order for the bores to be placed as close as possible to the racetracks coils inner radius, thereby maximizing the MFOM.

The cold mass is fixed to the central post of the cryostat. The cold mass supports are made of four G10 feet, connecting the reinforced bottom keystone box (referred to as KSB8) to the central part of the cryostat and transfer the weight load of the cold mass to the cryostat. KSB8 also provides a thermal property: the cold mass supports are not directly attached to the coils casings, thereby reducing the heat load on the windings and affecting less the stability of the magnet. The support feet are thermally connected to the thermal shield with copper braids, further reducing the heat load on KSB8. Moreover, KSB8 can be directly cooled to ensure that the magnet stability margins remain at the desired level.

Mechanical stops, which counteract forces along a specific axis, will be introduced at both ends of the cold mass to reduce the stress on the fixed support feet when the magnet is positioned at different inclination angles.


Searching for solar axions, the IAXO detectors need to track the sun for the longest possible period in order to increase the data-taking efficiency. Thus, the magnet needs to be rotated both horizontally and vertically by the largest possible angles. For vertical inclination a $\pm$ 25$^\circ$ movement is required (see later section \ref{sec:additional}, while the horizontal rotation should be stretched to a full 360$^\circ$ rotation before the magnet revolves back at a faster speed to its starting position.

The 250 tons magnet system will be supported at the center of mass of  the whole system through the cryostat central post (see Fig. \ref{fig:IAXO_sketch}), thus minimizing the torques acting on the support structure and allowing for simple rotation and inclination mechanisms. Accordingly, an altitude-over-azimuth mount configuration was chosen to support and rotate the magnet system, as described in section~\ref{sec:platform}. This mechanically simple mount, commonly used for very large telescopes, allows to separate vertical and horizontal rotations. The vertical movement is performed by two semi-circular structures (C-rings) with extension sections which are attached to the central part of the vacuum vessel. The C-rings distribute support forces from the rigid central part of the vessel to the C-rings pedestals, equipped with hydraulic elevation pads and drives. The pedestals are mounted on top of a 6.5~m high structural steel support frame which is situated on a wide rotating structural steel disk. The rotation of the disk is generated by a set of roller drives on a circular rail system.

The required magnet services, providing vacuum, helium supply, current and controls, are placed on top of the disk to couple their position to the horizontal rotation of the magnet. The magnet services are connected via a turret aligned with the rotation axis, thus simplifying the flexible cables and transfer lines arrangement. A set of flexible chains are guiding the services lines and cables from the different services boxes to the stationary connection point.

\subsection{Cryogenics}

The coil windings are cooled by conduction at a temperature of 4.5~K. The conceptual design of the cryogenic system is based on cooling with a forced flow of sub-cooled liquid helium at supercritical pressure. This avoids two-phase flow within the magnet cryostat and hence the complexity of controlling such a flow within a system whose inclination angle is continuously changing. The coolant flows in a piping system attached to the coil casings, allowing for conduction cooling in a manner similar to the ATLAS toroids \cite{tenKate:1158687,tenKate:1169275}.

The heat load on the magnet by radiation and conduction is $\sim$150~W at 4.5~K. In addition the thermal shield heat load is $\sim$1.6~kW. An acceptable thermal design goal is to limit the temperature rise in the coils to 0.1~K above the coolant temperature under the given heat loads.

\begin{figure}[!b]
\begin{center}
    \includegraphics[scale=0.42] {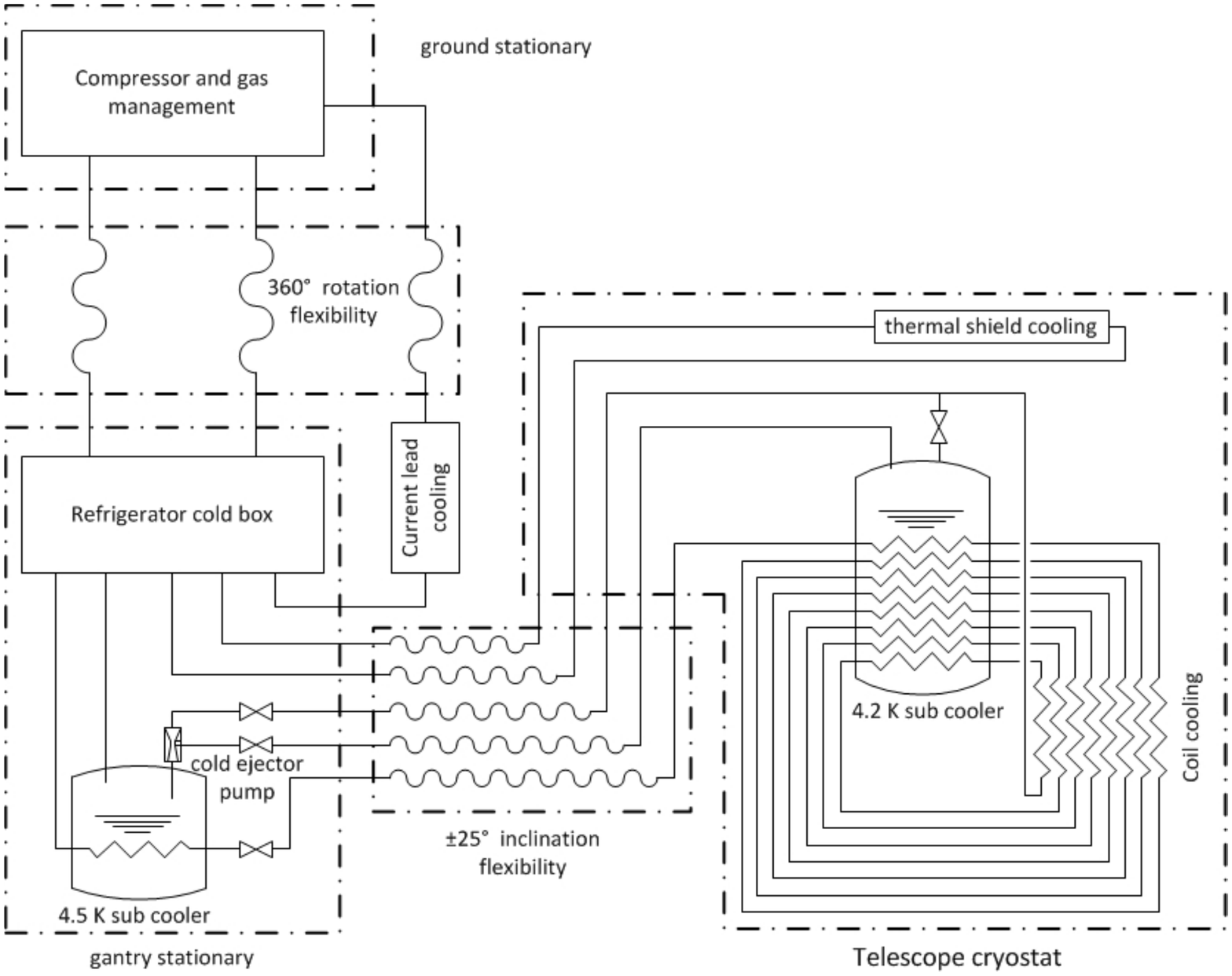}
    \caption{Flow diagram of the cryogenic system of the IAXO magnet.}
    \label{cryo}
\end{center}
\end{figure}

Fig.~\ref{cryo} shows a schematic flow diagram of the cryogenic system concept. It features the helium compression and gas management that is ground-stationary. The refrigerator cold box, current leads cryostat and a 4.5~K helium bath are integrated on the rotating disk that carries the structure of the helioscope. A helium bath operating at 4.2~K is connected to the magnet cryostat to follow its movement.

The magnet coils are cooled by a helium flow of 23~g/s, supplied at about 300~kPa and 4.6~K and sub-cooled in the 4.5~K bath. Before entering the cooling circuit of the first coil, the flow is cooled to 4.3~K in the 4.2~K bath. After passing through the cooling channels of each coil, the helium, then below 4.5~K, is re-cooled in the sub-cooler of the magnet cryostat. As the flow returns from the pipes of the last magnet coil, part of the helium is used to supply the 4.2~K sub-cooler and the remaining gas supplies the 4.5~K bath on the rotating disk. The latter flow is also used as a drive flow for a cold injector pump that pumps the 4.2~K sub-cooler at ambient pressure.

The thermal shield is cooled by a flow of 16~g/s gas at 16~bar between 40~K and 80~K. The cooling of the current leads is supplied at 20~K and 1.2~bar, which corresponds to the cooling used for the HTS current leads of the LHC machine~\cite{Ballarino:2007kx}. The path of the superconducting cables to the magnets is not shown in Fig.~\ref{cryo} but they could for example be integrated in the helium supply line.

The total equivalent capacity of the refrigerator results in a 360~W isothermal load at 4.5~K. Thus, the refrigerator cold box will be compact enough to be integrated together with the cryostat of the current leads on the gantry that is rotating with the helioscope. All cryogenic lines between the refrigerator and the magnet cryostat will therefore only need to compensate for the $\pm$ 25$^\circ$ inclination, and not for the 360$^\circ$ rotation that will be followed only by ambient temperature lines.

\subsection{Magnet system reliability and fault scenarios}

The IAXO magnet system is a complex combination of subsystems which work in harmony. Therefore, the anticipation of fault scenarios and the basic operational strategy in case of such failures should be dealt with already at the design stage. Here, we identify and describe the major fault cases which could interrupt the normal operation of the system:

\begin{itemize}

\item \textit{Cryogen leak:} Minor leaks in the cryogenic pipes will result in exceeding the vacuum system trip limits. In this case the safety system will initiate a slow magnet discharge. In the case of a rupture in the cryogen lines a rapid pressure rise in the vessel will occur. The vessel will remain protected by a set of relief valves, while a fast shutdown of the system will be initiated.

\item \textit{Vacuum failure:} The vacuum system is supported by safety valves, thus considerably decreasing the probability of a catastrophic vacuum failure . Normal vacuum system faults will be dealt with by hard-wired interlocks.

\item \textit{Quench protection system failure:} Total failure of the quench detection system or the heaters system will be avoided by using multiple detectors and heaters to give redundancy to the system. Nonetheless, the coils and conductors are designed to stand even such fault conditions so that a complete quench system failure will not lead to coil nor conductor damage.

\item \textit{Power failure:} If the mains power will fail to supply the magnet control systems, the supply will be secured by a uninterruptible power source (UPS). Nonetheless, such a fault scenario will initiate a slow discharge of the magnet.

\item \textit{Refrigeration supply failure:} A failure to supply cooling power from the refrigerator cold box to the current leads and bus-bars, thermal shield or the 4.2~K sub-cooler will initiate a slow discharge process.

\item \textit{Water supply failure:} Water supply failure to the power converter, vacuum pumps, etc. will result in a slow discharge.

\item \textit{Air supply failure:} Air supply is required for the steady operation of the vacuum system and the cryogenic system. A deficient air supply to these systems will lead to a slow discharge of the magnet.

\item \textit{Seismic disturbances:} The structure and movement system must sustain an additional sidewards load of at least 1.2g, which may be caused by a moderate seismic activity.

\end{itemize}

System reliability is an important issue when designing a complex assembly of subsystems such as the IAXO magnet system, let alone when the system is required to operate for long periods of time without exterior interference. Some key factors are to be noted when defining the magnet system's reliability: A fast discharge of the magnet should be initiated only when the magnet, experiment as a whole or personal safety is in danger. In all other cases a slow energy dump should take place. A UPS unit will maintain key services in order to enable a safe and controlled slow discharge in extreme cases. Lastly, routine maintenance is essential to avoid false magnet discharges.

\subsection{Cryostat assembly procedure and integration}

The IAXO detector will be placed in a light and confined structure, such as a dome or a framed tent, which will serve as the main site for the experiment. For this reason the assembly requires a hall with enough space to allow for the large tooling and infrastructure needed for the final cold mass and cryostat integration.

The assembly of the cold mass and the cryostat will be performed in five main steps: First, each of the eight warm bores, surrounded by 30 layers of super-insulation and a thermal shield, will be connected to one keystone box and two keystone plates to form eight sub-units. These sub-units will be attached, together with the coils casings, to the cold mass central support cylinder in order to assemble the complete cold mass. Cooling circuits will be installed and bonded to the surface of the coils casings already during fabrication. Additional cooling pipes will be attached to the cold mass when the latter is assembled. Next, the complete cold mass will be connected to the central part of the cryostat, where the cylindrical cold mass G10 based supports will be inserted into their their designated slots. The two cylindrical parts of the vessel will be connected to the central part, followed by the enclosure of the cryostat by the two Kl\"opper end plates which will be connected to the end flanges on both sides of the cryostat cylinders and to the bores. Lastly, the magnet vessel will be transferred to the main site where it will be attached to the movement system. The installation of services lines to the services turret, as well as the integration of the magnet system with the rest of the experiment's systems, will be performed at the last stage of system integration in the main site.

\subsection{T0 prototype coil for design validation and risk mitigation}
\label{sec:t0}

\begin{figure}[!t]
\label{t0}
\begin{center}
    \includegraphics[width=9cm] {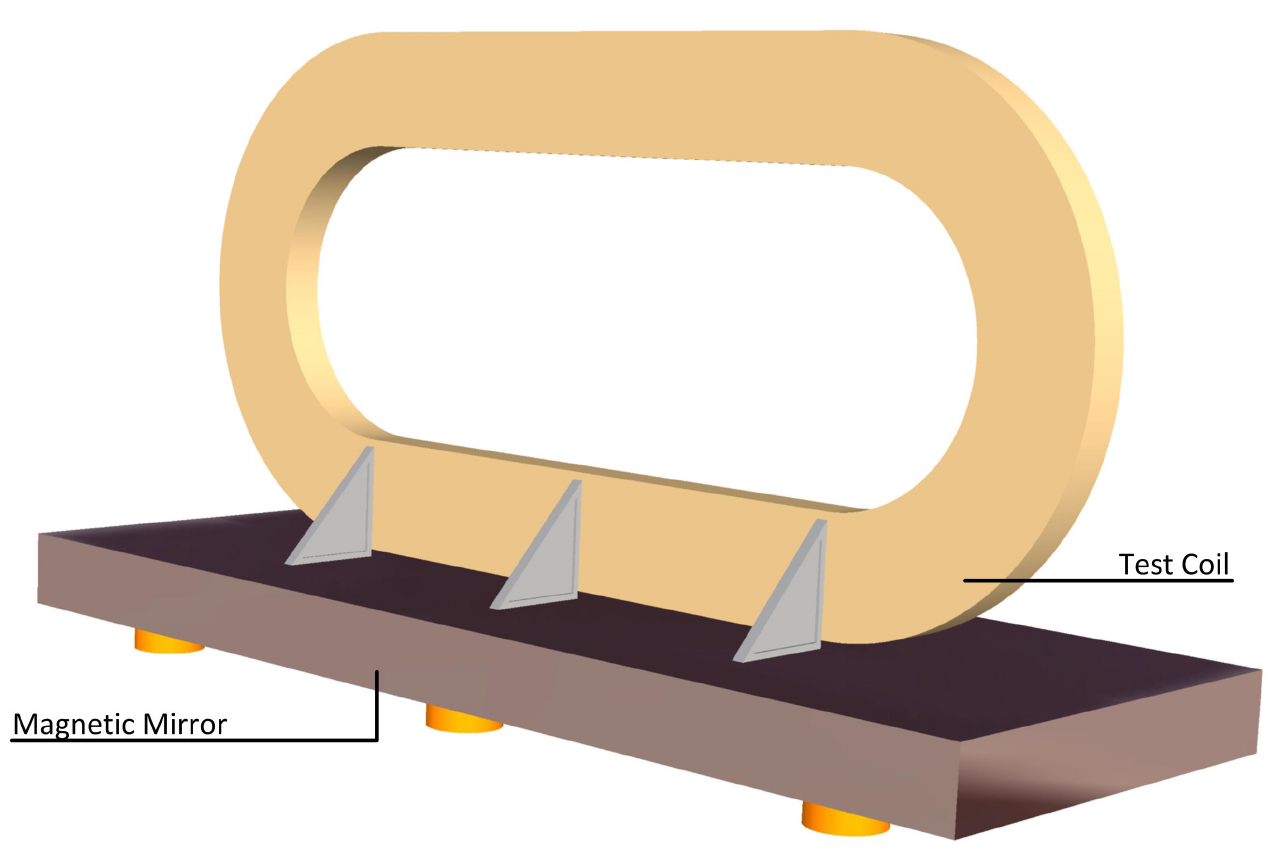}
    \caption{Possible arrangement of the two-meter long prototype coil T0 in combination with a cold iron yoke to generate the windings internal stress and force levels as in the full size detector toroid.}
    \label{fig:t0}
\end{center}
\end{figure}

Though the design of the toroid is based on the experience gained on the ATLAS toroids, still the IAXO toroid features a peak magnetic field of 5.4~T which is not trivial in terms of superconductor development and training behavior of the coil. In order to validate the design and thereby gain essential manufacturing experience that will flow back to the final manufacturing design, it is highly recommended to construct and test a single short prototype coil, called T0. This coil features the same windings cross section and cold mass design as for the full toroid, see Table~\ref{table1}, but its length is limited to two meters for reducing cost and enabling easy performance testing in an existing test facility.
A demonstration program for the T0 coil comprises:

\begin{itemize}

\item development and procurement of a 200~meter test length of Al stabilized NbTi/Cu conductor, extensive conductor qualification tests, followed by production of the 2x600~meter long units required for the T0 coil;

\item coil winding, coil casing production and cold mass integration;

\item and finally the performance test of the coil.

\end{itemize}

Ideally, during the test of the single shorter T0 coil the actual stress and Lorentz forces per meter as present in the full size toroid coil windings should be approximated in order to qualify the mechanical soundness of the coil windings and check its vulnerability for training. This can be achieved by testing the prototype coil at an excess operating current, eventually in combination with a cold-iron mirror temporarily attached to the test coil for this purpose. A test setup of the T0 coil adapted to the constraints of the test facility is shown in Fig.~\ref{fig:t0}.
With iron present this arrangement can produce the 50~MPa coil windings stress as in the full toroid at a test current of 14.9~kA (excess of 2.6~kA) and a peak magnetic field of 5.5~T. The pulling force on the straight section of the coil is then some 0.45~MN/m. Without iron a test current of 16.2~kA (excess of 3.9~kA) is needed for generating 50~MPa with 5.6~T peak magnetic field. However, in this case there is no pulling force on the coil inner beam.

\section{The IAXO x-ray optics}
\label{sec:optics}
\subsection{Basic considerations}

The purpose of the x-ray optics is to focus the putative x-ray signal to as small a spot as possible, and in doing so, reduce the size of the detector required and, ultimately, the detector background.   The performance of an x-ray optic is generally characterized by three basic properties:  the point spread function (PSF), the shape of the resultant spot; the throughput, $\epsilon_{o}$, the amount of incident photons properly focused by the optic; and the field-of-view (FOV), the extent to which the optic can focus off-axis photons.

Although x-ray optics can rely on refraction, diffraction or reflection, the large entrance pupil and energy band required for IAXO lead us to only consider grazing-incidence reflective optics.  To achieve the smallest spot $a$, the optics should have as short a focal length, $f$, as possible since the spot area grows quadratically with focal length, $a \propto f^{2}$.
At the same time, the individual mirrors that comprise the optic should have the highest x-ray reflectivity.  Reflectivity increases with decreasing graze angle, $\alpha$, and since $f \propto \frac{1}{\alpha}$, to achieve the highest throughput the optics should have as long a focal length as possible.

Further complicating the optical design is that the $\epsilon_o$, FOV and PSF of an optic have a complex dependence on the the incident photon energy $E$ and $\alpha$.

 \begin{itemize}
   \item  {\bf  Throughput:} there are many choices for the coatings of an x-ray mirror.  These coatings can have abrupt changes in reflectivity as a function of energy when the pass-band includes the characteristic absorption edges of the constituent materials.  And as already discussed, the reflectivity will be higher with decreasing graze angle.
   \item   {\bf  Field-of-view:}  The FOV is impacted by a phenomenon referred to a ``vignetting,'' the loss of photons that pass through the entrance aperture of the optic but are not properly focused on the focal plane. Vignetting is more severe at lower graze angles and increases with the off-axis position.  Vignetting is a geometric effect and would occur even if the coatings have 100\% reflectivity.   When realistic coatings, with their own dependence on $E$, are accounted for, the FOV becomes dependent on the photon energy and decreases at higher energies.
   \item     {\bf Point Spread Function:}  The PSF of an x-ray telescope depends on several factors including the basic design, the long spatial frequency errors (usually referred to as figure errors) and short spatial frequency errors (usually refereed to as finish errors).  These first two factors can be accounted for using geometric optics treatments and do not have a formal energy dependence.  But like the FOV, once realistic coatings are considered, the PSF can take on a mild energy dependence.  Finish errors can be accounted for using wave optics treatments (e.g., scattering theory) that depend on both $E$ and $\alpha$.  Several authors have shown that the transition between the valid use of geometric and wave optics itself has a dependence on $E$ and $\alpha$, so the final energy dependence of the PSF is not easily described by a simple relationship.
 \end{itemize}

There are two basic families of reflective x-ray optics:  those that employ two reflections to (nearly) satisfy the Abbe sine rule and have excellent imaging properties across its FOV; those that employ a single reflection and have poor imaging properties.  The former include a family of designs originally proposed by Wolter and include telescopes and point-to-point imagers; the later include concentrators and collimators.

Since the x-rays produced via the conversion of axions to photons in the IAXO magnet have the same directionality of the axions, the optic need only have a FOV slightly larger than the inner 3 arcminute ($\sim$0.9 mrad) Solar disk, the region of axion production.   Moreover, the fact the emission is from a uniformly filled extended region means that, to first order, a telescope or concentrator with the same focal length $f$ will result in the same focused spot of $\sim 1.0 \times f_{\rm m}$~mm, where $f_{\rm m}$ is the focal length in meters.
The focal length of the optic, be it a collimator or a telescope, depends on the radius of the largest shell and maximum graze angle that can still result in a high reflectivity from the mirror shell.  For a telescope, the relationship is $f \propto \frac{\rho_{\rm max}}{4 \alpha}$, while for a collimator it is $f \propto \frac{\rho_{\rm max}}{2 \alpha}$.

To zeroth order, the x-ray reflectivity at a single energy of a single metal film is near-unity up to a certain angle, called the critical angle, and then zero above that angle.  If this were strictly true, a telescope would be the clear winning design for IAXO, since it would have half the focal length, and hence half the spot diameter and one-fourth the area of a collimator.

However, we know that the reflectivity in the 1$-$10~keV has a more complex relationship as a function of energy and $\alpha$.
The throughput of the optics will depend on the reflectivity, which in turn depends on the coating material and graze angle, and the optical design, which determines the number of reflections a photon will experience as it passes through the optic:  for a collimator, $n=1$ while for a telescope, $n=2$.

\subsection{Fabrication techniques for reflective optics}

The x-ray astronomy community has designed, built and flown x-ray telescopes on more than ten satellite missions, and they have developed a number of techniques for fabricating the telescopes.  For each technology, we give
a brief description and cite examples of telescopes
that rely on it.  Broadly speaking, telescopes can be classed
into two groups that depend on how they are assembled.
Segmented optics rely on several individual
pieces of substrates to complete a single layer.  (The appropriate
analogy is the way a barrel is assembled from many individual staves.)
Integral-shell optics are just that:  the hyperbolic or parabolic
shell is a single monolithic piece.

\subsubsection{Segmented optics:  rolled aluminum substrates}
Telescopes formed from segmented aluminum substrates were first
utilized for the broad band x-ray telescope (BBXRT) that flew on
the Space Shuttle in 1990~\cite{ps85}.   Later missions that used the
same approach included:  {\it ASCA}~\cite{sjs+95}, launched in 1993;
 SODART~\cite{cmh+97}, completed in 1995 but never launched; InFoc$\mu$s~\cite{infocus}, a hard x-ray balloon-borne instrument flown in 2004; and {\it Suzaku}~\cite{ssc+07}, launched in 2005.
Aluminum substrates will also be used for the
soft and hard X-ray telescopes on the
upcoming JAXA Astro-H (also called NeXT) mission, scheduled for launch in 2014~\cite{okm+08}.

\subsubsection{Segmented optics:  glass substrates}
Although using glass substrates for an x-ray telescope was explored as far as back
as the 1980s \cite{labov88}, it was not fully realized until 2005 with
the launch of HEFT~\cite{hcch+05}.  HEFT had three, hard x-ray telescopes, each consisting of as many
as 72 layers.  HEFT was the pathfinder for NASA's {\it NuSTAR}~\cite{nustar2013}, launched in 2012 and the first satellite mission to use focusing x-ray optics to image in the hard x-ray band up to 80~keV. Each of {\it NuSTAR's} two telescope consists of 130 layers, comprised of more than 2300 multilayer-coated pieces of glass.    Finally, slumped glass is a candidate technology being developed by several groups for future NASA and ESA missions, like ATHENA (see, e.g., \cite{wvf10} and \cite{zab+10} ).

\subsubsection{Segmented optics:  silicon substrates}
Another technology being pursued for ATHENA are silicon pore optics~\cite{cga+10}, which
consists of silicon wafers that have a reflective coating on one
side and etched support structures on the other.  Individual segments
are stacked on top of each other to build nested layers.  Prototype
optics have been built and tested, but there are no operational x-ray
telescope yet to use this method.

\subsubsection{Integral shell optics:  replication}
Replicated optics are created by growing
the mirror,  usually a nickel-based alloy,
on top of a precisely figured and polished mandrel or
master.  The completely-formed shell is separated from
the mandrel, and two unique mandrels are required for each individual layer (one for the parabolic-shaped primary, another for the hyperbolic-shaped secondary).
Missions that have utilized replicated x-ray telescopes include:
{\it XMM}~\cite{jla+01}, launched in 1999; {\it Beppo-SAX}~\cite{cbc+88}, launched in 1996; {\it ABRIXAS}~\cite{eak+98}, launched in 1999;
the balloon-borne HERO mission~\cite{heroXRAY}, first flown in 2002; and the sounding rocket mission FOXSI, currently under development.
It is important to mention that CAST currently employs a flight-spare telescope from {\it ABRIXAS}.

\subsubsection{Integral shell optics:  monolithic glass}
For completeness, we mention telescopes formed from
monolithic pieces of glass.  Although these telescopes have excellent focusing quality and have produced some of the best images of the x-ray sky, because of the cost and weight, no future
mission is expected to use this approach. Missions that have utilized monolithic optics include:
{\it Einstein}~\cite{gbb+79}, launched in (1978); {\it RoSAT},~\cite{trumper83} launched in 1980;
and the {\it Chandra X-ray Observatory}~\cite{wtvo00}, launched in 1998.

\subsection{The baseline technology for IAXO}
For IAXO, we have adopted segmented, slumped glass optics as the baseline fabrication approach for several reasons.  First, the technology is mature and has been developed by members of the IAXO collaboration, most recently for the NuSTAR satellite mission.   Second, this approach easily facilitates the deposition of single-layer or multi-layer reflective coatings.  Third, it is the least expensive of the fabrication techniques.  Fourth, the imaging requirement for solar observations for IAXO is very modest--focusing the central 3 arcminute core of the Sun.  Although other optics technologies may have better  resolution than slumped glass, they would not produce a significantly smaller focused spot of the solar core.

\subsection{The IAXO x-ray telescopes}

\begin{figure}[!t]
\centering
\includegraphics[width=0.65\textwidth]{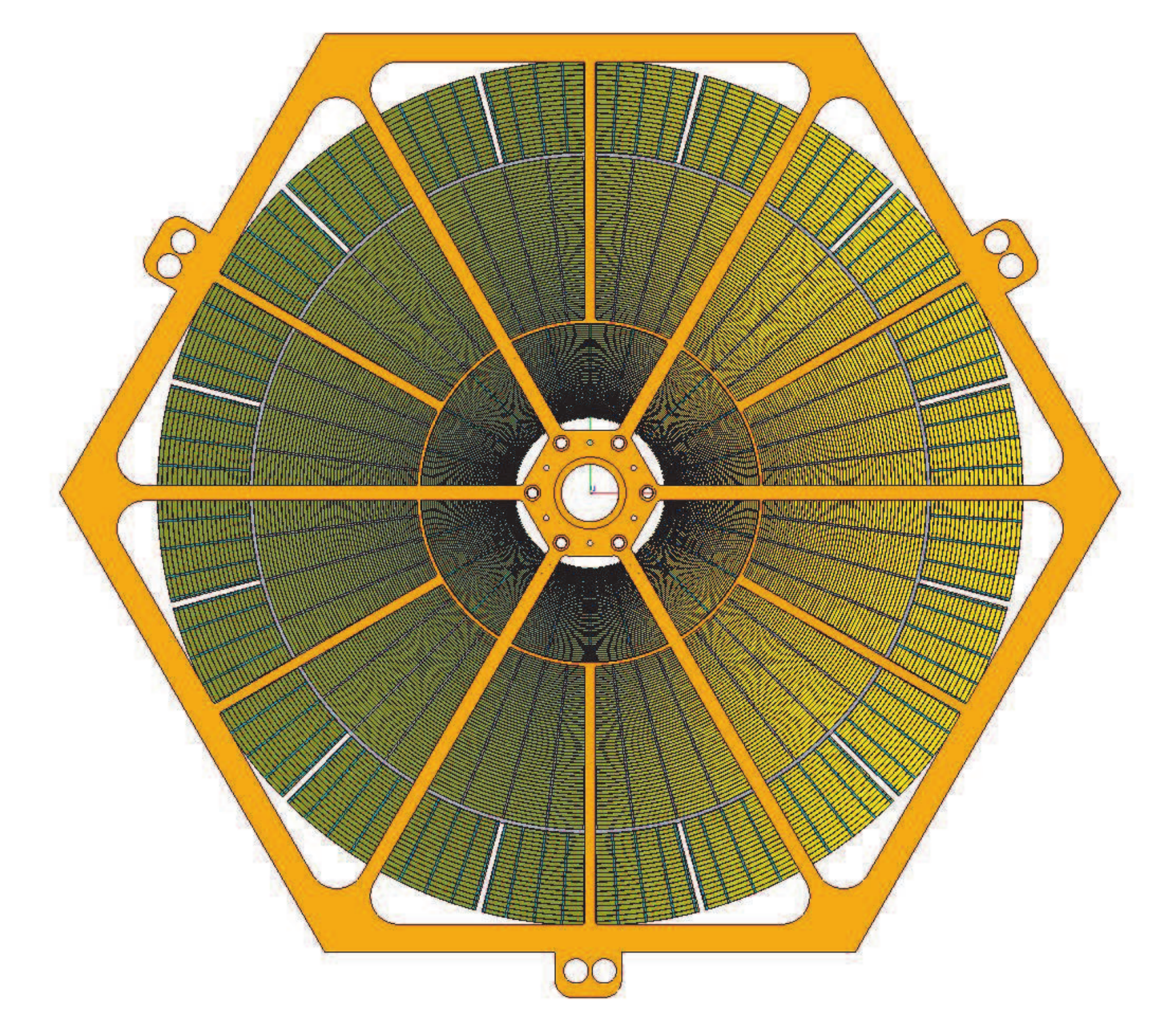}
\caption{\label{fig:optic1}An edge-on view of one IAXO optic, including the hexagonal ``spider'' structure that will be used to mount the optic into the magnet bores.  The thousands of individual mirror segments are visible.}
\end{figure}

\begin{figure}[!b]
\centering
\includegraphics[width=0.65\textwidth]{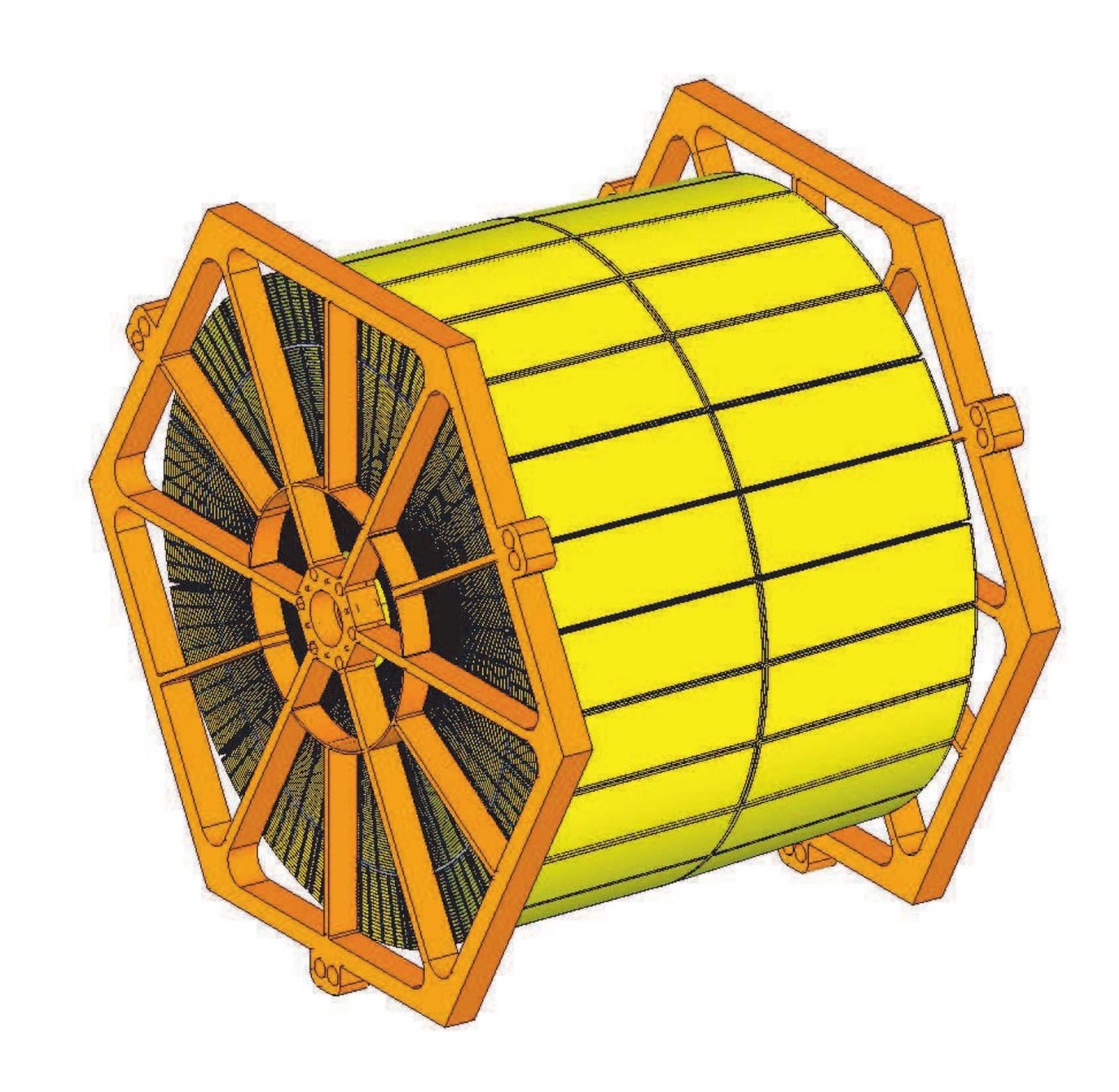}
\caption{\label{fig:optic2}An isomorphic side-view of the telescopes and the spider mounting structures.}
\end{figure}

\subsubsection{Design and optimization of the IAXO x-ray telescopes}

The optical prescription and reflective coatings were identified by a systematic search of a multi-dimensional parameter space that accounted for the detector efficiency, axion spectrum, optics properties and recipe of the reflective coatings. The total optics and detector figure of merit, $f_{DO}$ was then computed.  The  optical prescription and multilayer recipes presented below produced the highest $f_{DO}$.
It is important to note that the telescope optimization \textit{must} account for the axion spectrum and detector efficiency and cannot be performed independently.  If this process does not include these energy dependent terms, $f_{DO}$ will not achieve the highest possible value.

Telescope prescriptions were generated for designs that had a fixed maximum radius of 300 mm and a minimum radius of 50 mm, with the focal length varied between 4 and 10 m, in increments of 1 m.  As the focal length is increased and the graze angle, $\alpha$, decreases and the number of nested layers increases.  For example, the $f = 4$~m design has 110 nested layers, while the $f = 10$~m design has more than 230 layers.

\begin{figure}[!t]
\centering
\includegraphics[width=10cm]{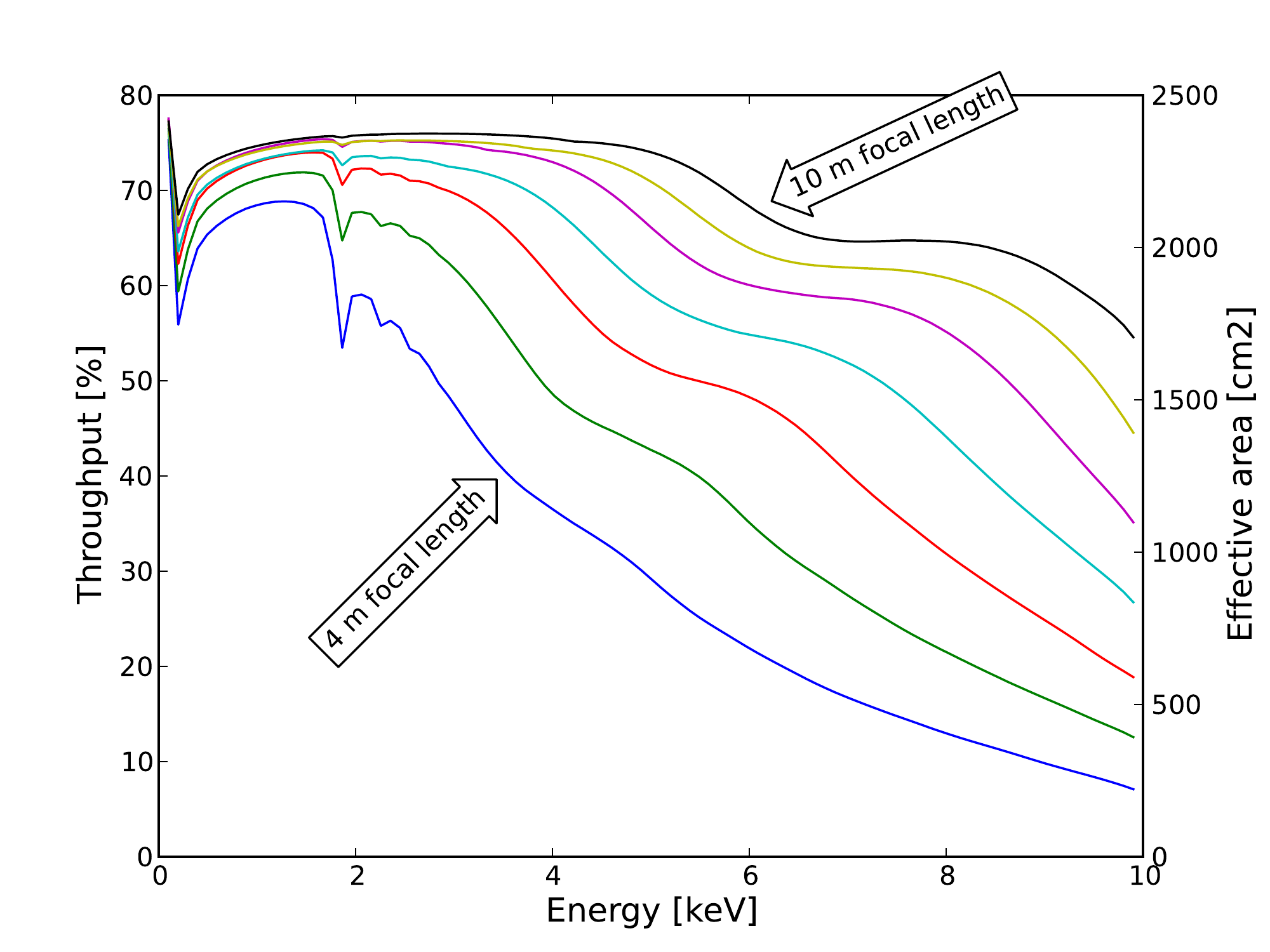}
\caption{\label{fig:opticsArea1}Effective area (right axis) and throughput/efficiency (left axis) versus photon energy for a single telescope for different focal lengths considered, from $f=4$~m (lowest curve) up to $f=10$~m (highest curve). Effective area grows as the focal length is increased.}
\end{figure}

Traditionally, x-ray telescopes have relied on single layer coatings of metals like Au or Ir to achieve high throughput in the 1$-$10~keV band.  More recently, missions designed for hard x-ray observations, like NuSTAR and ASTRO-H, have employed multilayer coatings to achieve high reflectivity up to $\sim$80~keV.  We explored combinations of both for IAXO.  Although it is theoretically possible to optimize the coating for each layer of the telescope, this would impose a high penalty in resources when depositing multilayers on the substrates.   Instead, we divided the layers into ten sub-groups, with each sub-group of layers receiving the same multilayer coating.  A similar strategy was successfully implemented for NuSTAR \cite{mhmc+09}, and this approach allowed the multilayer deposition tools to be used efficiently.

Material types investigated were single layers of W and W/B$_{4}$C multilayers. Other types/ combinations to consider are W/Si, Pt/B$_{4}$C, Ir/B$_{4}$C and Ni/B$_{4}$C. W/B$_{4}$C and W/Si are well understood coatings for x-ray reflectivity and considerably less expensive to use W than Pt or Ir. Using B$_{4}$C instead of Si as the light material will give increasedF reflectivity at 1$-$4~keV, but also gives slightly higher stress in the coating. Ni/B$_{4}$C coatings are not well understood and can give a high interfacial roughness between light and heavy material, but performs similar to W/B$_{4}$C and Ir/B$_{4}$C at 1$-$10~keV.


At a given substrate incident angle, $\alpha$, the coating geometry was optimized by trying every combination in a parameter space of $n$ (number of bilayers), $d_{\rm min}$  (minimum bilayer thickness), $d_{\rm max}$ (maximum bilayer thickness) and $\Gamma$ , the ratio between the thickness of the heavy material with respect to the total thickness of the bilayer.  For every combination, the x-ray reflectivity was calculated using IMD~\cite{windt98}.
One of the basic properties of any x-ray telescope is the effective area,
{\rm EA}, the energy-dependent effective aperture of the telescope that accounts for finite reflectivity of individual mirror elements and physical obscuration present in the telescope (e.g., from the support structures used to fabricate the optics and the finite thickness of the substrate which absorbs incoming photons). The effective area of an individual layer $i$ is given by:
\begin{equation}
{\rm EA}(E)_{i} = {\rm GA}_{i} \times R_{i}(E,\alpha)^{2} \times 0.8,
\end{equation}
where ${\rm GA}_{i}$ is the projected geometric area of the individual layer $i$, $R_{i}(E,\alpha)$ is the reflectivity of the coatings on layer $i$ and the constant factor of 0.8 accounts for obscuration.  The total area is given by:

\begin{equation}
{\rm EA}(E) = \sum_{i=1}^{N} {\rm EA}_{i}(E),
\end{equation}

\noindent where $N$ is the total number of layers. Figure~\ref{fig:opticsArea1} shows the expected behavior of the effective area increasing as the focal length grows. Again, this behavior arises from the fact that longer focal lengths results in shallower incident angles, and  reflectivity increases with decreasing graze angles.

\begin{figure}[!t]
\centering
\includegraphics[width=10cm]{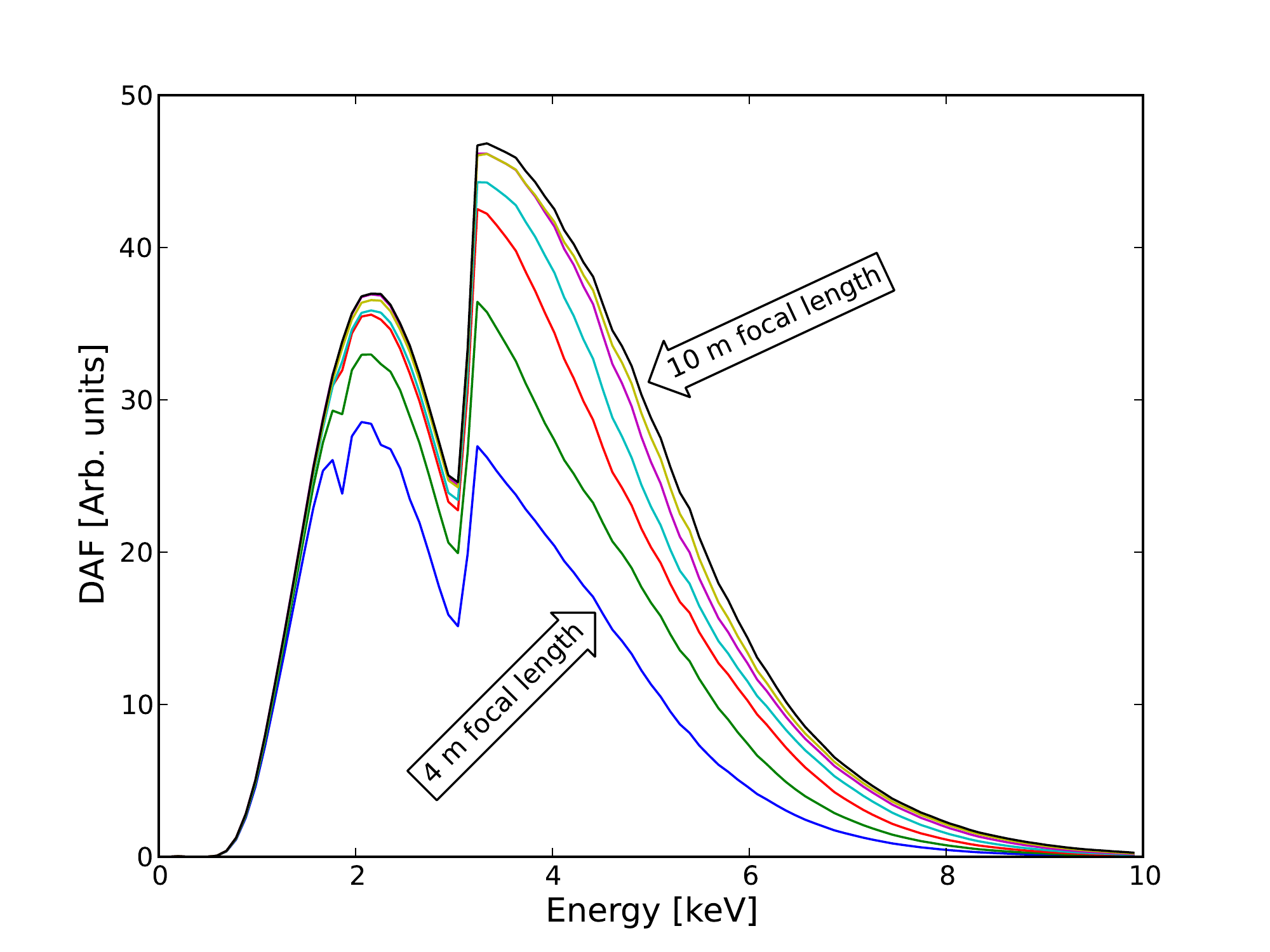}
\caption{\label{fig:opticsArea2}${\rm DAF}$ versus photon energy $E$ for a single telescope, and for the different focal lengths considered, from $f=4$~m (lowest curve) up to $f=10$~m (highest curve). The significant structure now present is due to absorption edges in detector and coating materials and the shape of the solar axion spectrum.
}
\end{figure}

The energy-dependent optics throughput or efficiency, $\epsilon_{o}(E)$, is simply the ${\rm EA}(E)$ divided by the geometric area of the entrance pupil:
\begin{equation}
\epsilon_{o}(E) = \frac{{\rm EA}(E) [{\rm m}^2]} {\pi (0.3^2 - 0.05^2) [{\rm m}^{2}]}
\end{equation}
The plot of Fig.~\ref{fig:opticsArea1} displays this quantity for different focal lengths.


%
In order to build a meaningful figure or merit we multiply the optics throughput by the energy-dependent axion flux $\frac{{d} \phi}{{d} E}$(E) expected from Primakoff production at the Sun~\cite{Irastorza:2011gs} and the detector efficiency $\epsilon_{d}(E)$. The resulting quantity, that we call ``detected axion flux'' (${\rm DAF}(E)$),

\begin{equation}
{\rm DAF}(E) =   \sum_{i=1}^{N} {\rm EA}_{i}(E) \times  \epsilon_{d}(E) \times \frac{{d} \phi}{{d} E}(E)
\end{equation}

\noindent is actually proportional to a hypothetical axion signal in IAXO, and is plotted in Fig.~\ref{fig:opticsArea2}.

\begin{figure}[!t]
\centering
\includegraphics[width=10cm]{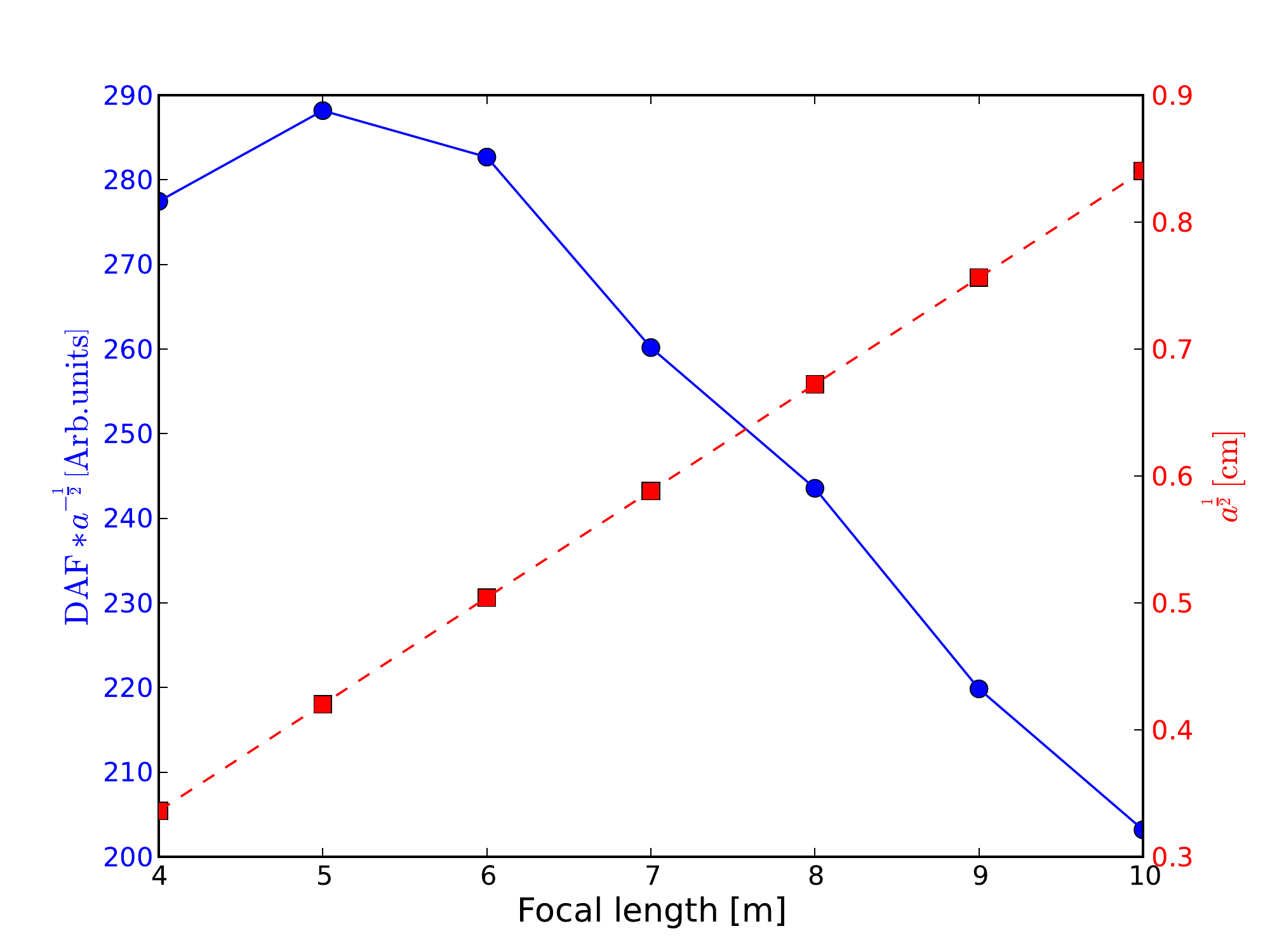}
\caption{\label{fig:opticsArea3}Value of the focal spot size $\sqrt{a}$ (red squares and dashed line, right axis) and the figure of merit $f_O$
(blue circles and solid line, left axis) versus focal length $f$. The optimal figure of merit is found for $f=5$~m.
}
\end{figure}

To find the optimal focal length, we need to maximize the integral of the DAF from 1$-$10~keV divided by the square root of the spot size (one could see this as the contribution from the optics to the figure of merit $f_{DO}$ as defined in~\cite{Irastorza:2011gs}):
\begin{equation}
 f_{O} \equiv \int_{E=1~keV}^{10~keV} \left( \frac{{\rm DAF}(E)}{\sqrt{a}} \right) dE.
\label{eqn:opticsFOM}
\end{equation}

The only quantity left to compute is the spot-size, $a$. The point-spread-function (PSF) of any x-ray telescope has a complex shape, and the spot-size is computed by first taking the integral of the PSF to compute the encircled energy function (EEF), a measure of how much focused x-ray light is contained within the diameter of a particular size.   For example, a common measure of the focusing quality of an x-ray telescope is to determine the  50\% value of the EEF, that is to determine the smallest diameter extraction region that contains 50\% of the power. This is often referred to as the half-power diameter or HPD.

The spot-size will depend on both the physical size of the object imaged, in this case the 3 arcminute (0.87 mrad) central core of the Sun, and the intrinsic imaging capability of the x-ray optic, i.e. the size of the resultant spot when the telescope images a point-like source.   To first order, then, the overall spot size $s_{total}$, measured in angular extent,  will be the root mean square of the object size $s_{obj}$ and the optic quality $s_{opt}$:
\begin{equation}
s_{total} = \sqrt{s^{2}_{obj}  + s^{2}_{opt} }
\end{equation}
Based on the performance of the NuSTAR x-ray telescopes~\cite{nustar2013}, we assume for the nominal design of the telescopes a HPD of 1~arcmin (0.29 mrad) and an 80\% EEF of 2~arcmin (0.58 mrad).
The angular spot size then becomes:
\begin{equation}
s_{total} = \sqrt{s^{2}_{obj}  + s^{2}_{opt}} =  \sqrt{ {0.87}^2  + 0.58^{2}} = 1.0~{\rm mrad}
\end{equation}
As discussed above, the spatial diameter of the spot is simply $f \times s_{total}$ and the spot area becomes:
\begin{equation}
a= \frac{\pi}{4} \left( s_{total} \times f \right)^2
\end{equation}

\subsubsection{Properties of the IAXO x-ray telescopes}

Fig.~\ref{fig:opticsArea3} shows $\sqrt{a}$ as well as $f_{O}$, as calculated in Eq.~\ref{eqn:opticsFOM}, as function of the focal length.
The optimal focal length is found to be $f = 5$~m. 
This parameter and the considerations exposed in previous sections fix the design proposed of the IAXO optics. Different engineering drawings of the optics are shown in Fig.~\ref{fig:optic1} and \ref{fig:optic2}, where the 123 nested layers can be seen. Finally, its main design parameters are listed in Table~\ref{tab:optics}.


\begin{table}[!t]
\begin{center}
\begin{tabular}{lr}
  \hline
  Telescopes & 8 \\
  $N$, Layers (or shells) per telescope & 123 \\
  Segments per telescope & 2172 \\
  Geometric area of glass per telescope & 0.38~m$^{2}$ \\
  Focal length & 5.0~m \\
  Inner radius & 50~mm \\
  Outer Radius & 300~mm \\
  Minimum graze angle & 2.63 mrad \\
  Maximum graze angle & 15.0 mrad \\
  Coatings & W/B$_{4}$C multilayers \\
  Pass band & 1$-$10~keV \\
  IAXO Nominal, 50\% EEF (HPD) & 0.29 mrad \\
  IAXO Enhanced, 50\% EEF (HPD) & 0.23 mrad \\
  IAXO Nominal, 80\% EEF &  0.58 mrad \\
  IAXO Enhanced, 90\% EEF & 0.58 mrad \\
  FOV & 2.9 mrad\\
  \hline
\end{tabular}
\end{center}
\caption{Main design parameters of the IAXO x-ray telescopes.}
\label{tab:optics}
\end{table}

\subsection{Final considerations}
Our preliminary scoping study has made simplifying assumptions that will be revisited for the final design study.
\begin{itemize}
\item We have assumed the axion spectrum and intensity is uniformly emitted from a region 3 arcmin in extent.  We must include the actual distributions in a full Monte Carlo model of the system performance.
\item We have computed effective area for an on-axis point source.   When the solar extent is included in ray-tracing, the area will decrease by a small amount.
\item We have not accounted for non-specular scattering.
\item We have assumed the encircled energy function (EEF) evaluated at 50\% (i.e., the half-power-diameter) is 1 arcminute and the EEF evaluated at 80\% is 2 arcminute.
\item We have only coarsely studied how the focal length $f$ influences the FOM in increments of 1~m.
\end{itemize}

\section{Ultra-low background x-ray detectors for IAXO}
\label{sec:detectors}
The baseline technology for the low background x-ray detectors for IAXO are small Time Projection Chambers (TPCs), with a thin window for the entrance of x-rays and a pixelated Micromegas readout, manufactured with the microbulk technique. This kind of detector has already been used in CAST, and has been the object of intense development in recent years, mainly within the T-REX R\&D project~\cite{Irastorza2011_EAS,Dafni:2012fi,Dafni:2012zz}, funded by the European Research Council (ERC). The CAST microbulk detectors have achieved record levels of background and, as described below, they offer the best prospects to meet the requirements for IAXO.

\subsection{State of the art}

\begin{figure}[b!]
\begin{center}
\includegraphics[height=65mm]{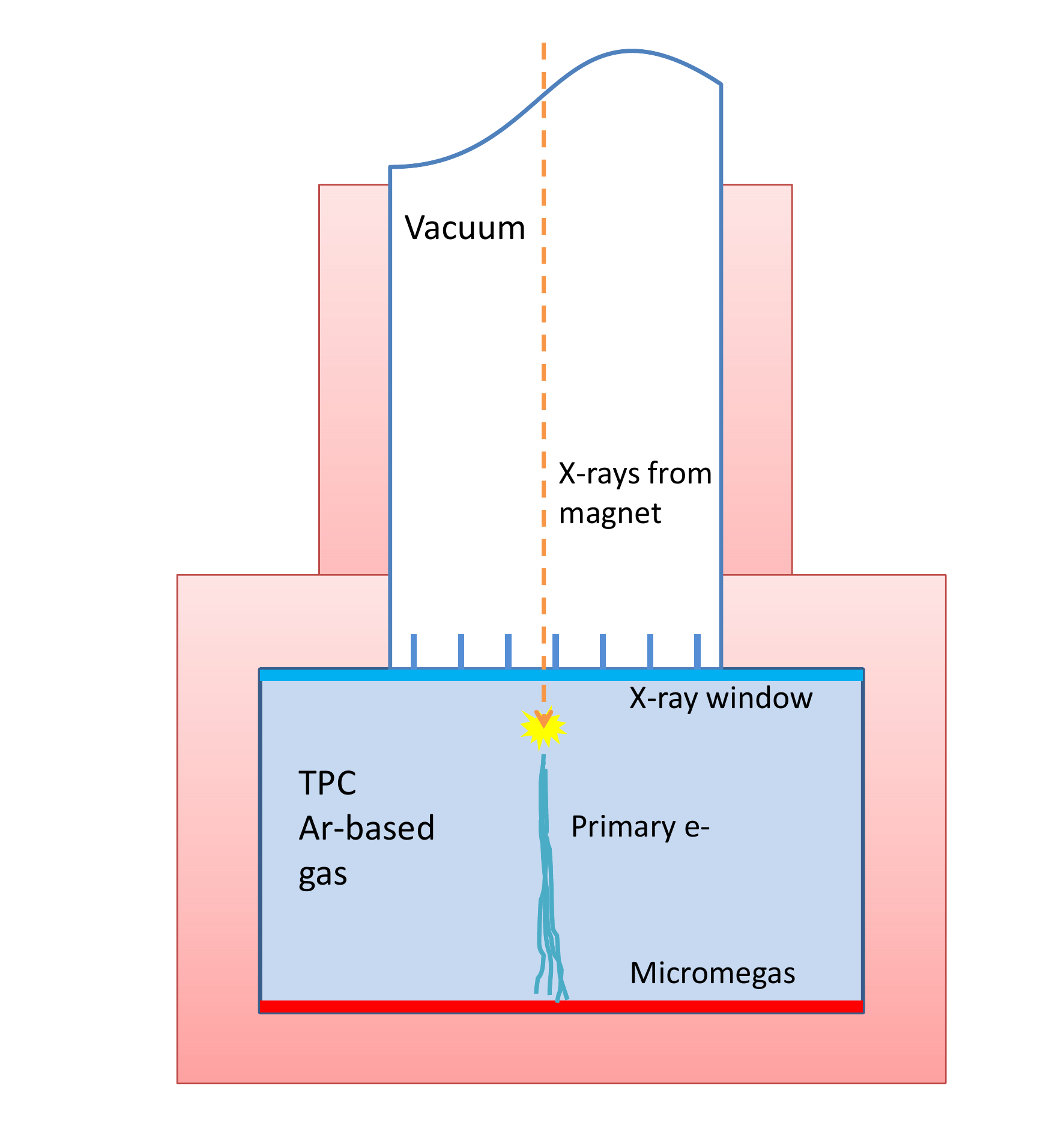}
\includegraphics[trim=1.5cm 0cm 1.5cm 0cm, clip=true,height=65mm]{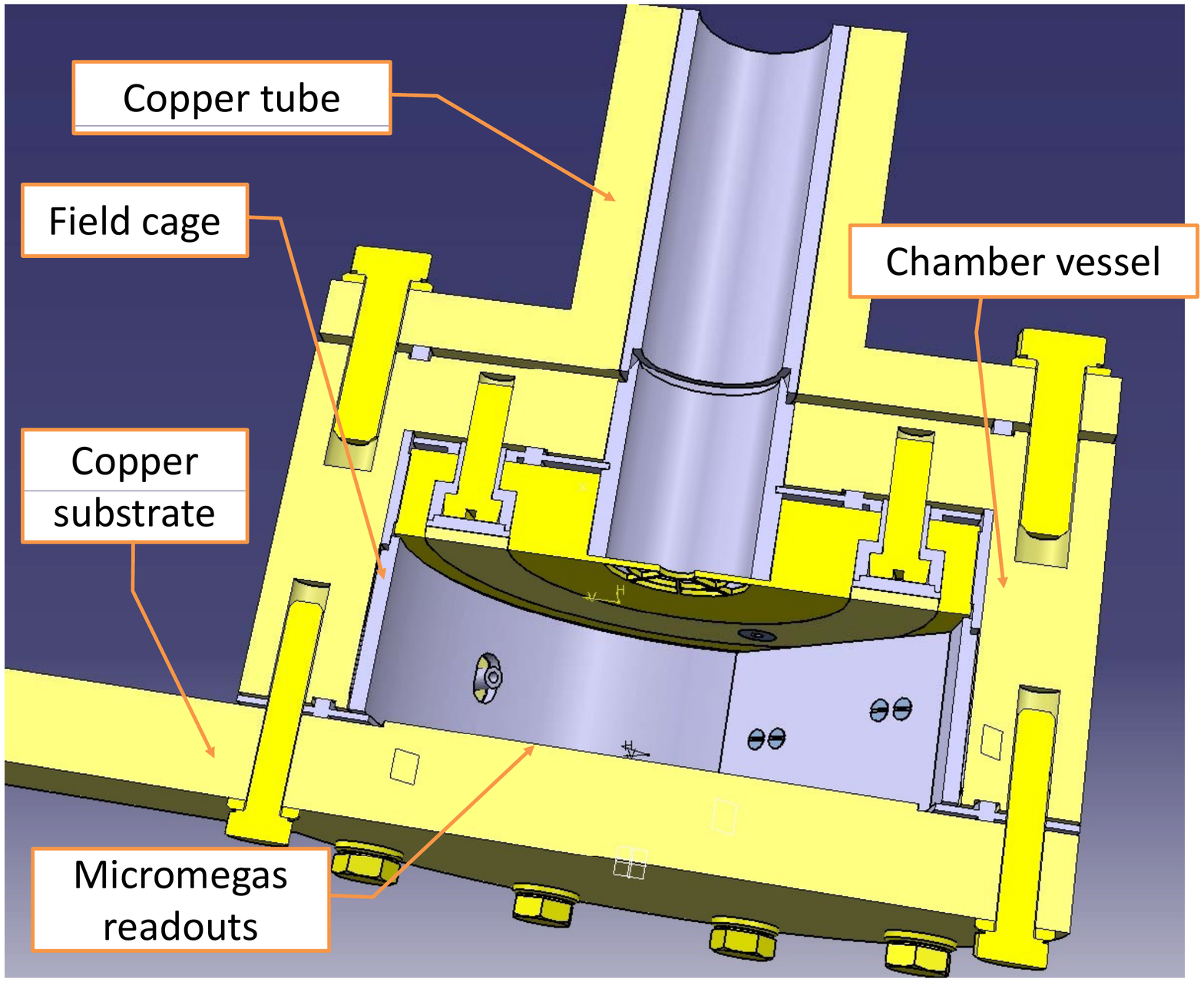}
\caption{\label{fig:det_concept} Left: Scheme of the detection principle of Micromegas detectors in IAXO. Right: Design of the IAXO detector prototype.}
\end{center}
\end{figure}


The detection concept is sketched on the left of figure~\ref{fig:det_concept}. The x-rays coming from the magnet enter the detector through a thin window (e.g. aluminized mylar), which is also the cathode of the TPC. This window holds the detector gas, so it must be sufficiently gas-tight and withstand the pressure difference, while being sufficiently transparent to the x-rays so as not to affect the efficiency of the detector. The drift distance $z$ of the TPC is adjusted so that the conversion volume contains enough gas to efficiently stop x-rays of the required energies. The design choice in CAST detectors has been $z=3$~cm at 1.5 bar of an argon gas mixture (usually Ar$-$2.3\% isobutane). The primary charge created by the interaction of x-rays drifts towards the anode of the TPC, where it is amplified by a Micromegas structure.

Micromegas readouts~\cite{Giomataris:1995fq,Giomataris:2004aa} make use of a metallic micromesh suspended over a
(usually pixellated) anode plane by means of insulator pillars, defining an amplification
gap of the order of 50 to 150 $\mu$m. Primary electrons go through
the micromesh holes and trigger an avalanche inside the gap, inducing detectable signals
both in the anode pixels and in the mesh. It is known \cite{Giomataris:2003pd} that the way the amplification
develops in a Micromegas gap is such that its gain $G$ is less dependent on geometrical
factors (the gap size) or environmental ones (like the temperature or pressure of the gas)
than conventional multiwire planes or other types of micropattern detectors based on
charge amplification. This fact allows in general for higher time stability and spatial
homogeneity in the response of Micromegas, in addition to better energy resolution.

\begin{figure}[b!]
\begin{center}
\includegraphics[height=9cm]{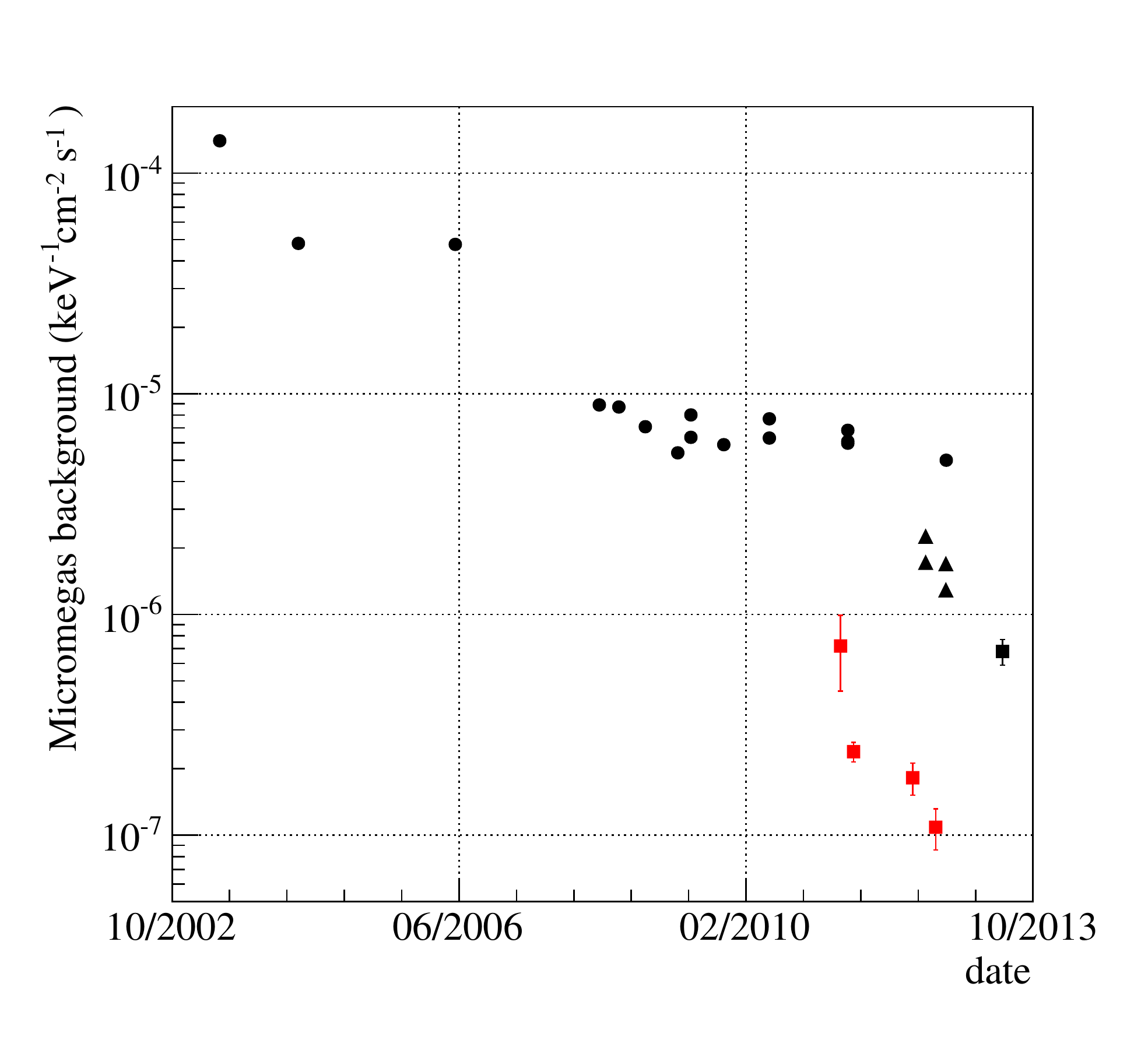}
\caption{\label{fig:plotestrella} Micromegas background history since the first detector installation in CAST in 2002. The black points correspond to the values obtained with the different detectors in the CAST hall. The black squared point is the last result obtained in the 2013 data taking campaign. Red points correspond to the values obtained with different shielding configuration in the Canfranc Underground Laboratory. }
\end{center}
\end{figure}

\hyphenation{rela-ti-ve}

These advantages together with the possibility of easily building large areas, its robustness, the relative cost-effectiveness and the high flexibility in patterning the anode plane, has spread the use of Micromegas in many areas of high energy physics. There are several fabrication techniques of Micromegas, but we focus on a recent one, \textit{microbulk }Micromegas~\cite{Andriamonje:2010zz}, originally developed at CERN and CEA. In these readouts the Micromegas amplification structures are obtained out of a double copper-clad kapton sheet, by chemically removing part of the kapton. This
technique is known to yield the highest precision in the gap homogeneity and, because of
that, the best energy resolutions among Micropattern detectors. Moreover, because of the raw materials, microbulk Micromegas are very radiopure objects~\cite{Cebrian:2010nw}, a desirable property in low background applications. Since the development of this type of Micromegas, these detectors have been studied, proposed or applied in an increasing number of low background applications~\cite{Irastorza2011_EAS}.
Indeed, the microbulk detectors have been used for the first time in a real data taking in CAST. In turn, CAST has been a test bench for this technology, that has evolved throughout the lifetime of the experiment.

\hyphenation{corres-pond}

\begin{figure}[b!]
\begin{center}
\includegraphics[height=5cm]{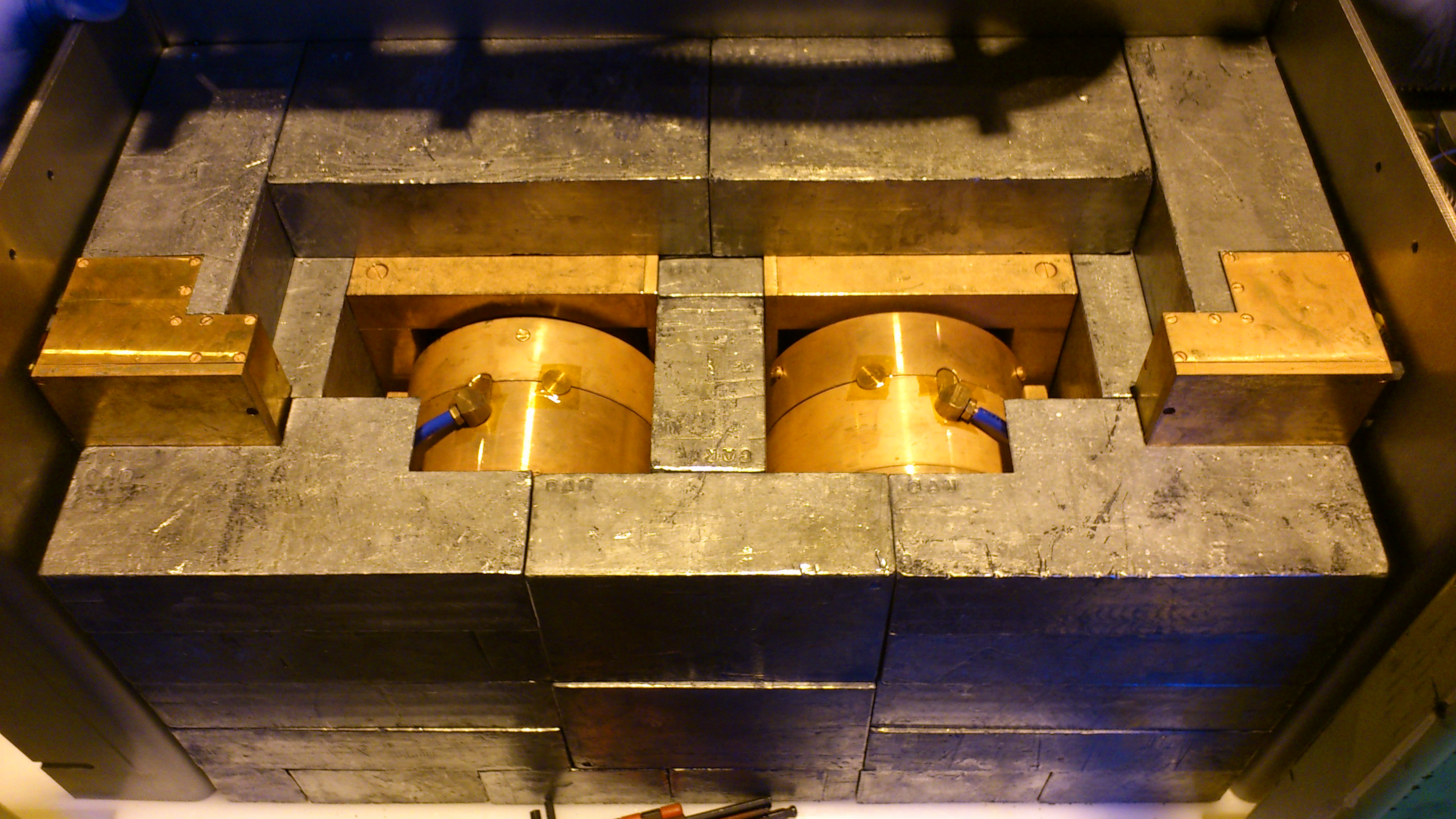}
\includegraphics[height=5cm]{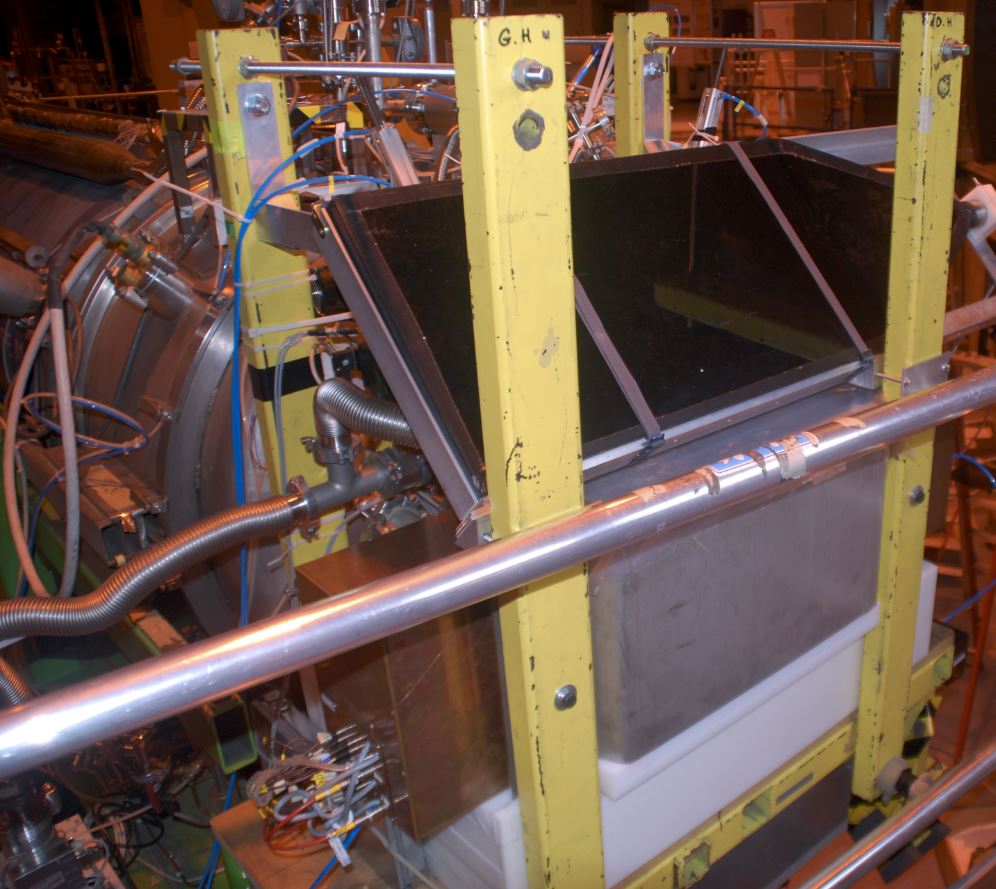}
\caption{\label{fig:shieldinginstallation} Left: picture of the CAST sunset side setup (incomplete), showing the inner copper shielding surrounded by pure lead pieces. Right: picture of the complete setup with the cosmic muon veto.}
\end{center}
\end{figure}

The plot of figure~\ref{fig:plotestrella} shows the Micromegas background history in CAST. The black points correspond to nominal background of data taking campaigns in the experiment, while red points correspond to the background level obtained in special runs in a test setup located underground, at the Laboratorio Subterr\'{a}neo de Canfranc (LSC). They always represent the average background in the energy region of interest 2$-$7 keV. The improvements obtained over time are due to continuous development of the detector setups. These improvements regard the following aspects:


\begin{itemize}
\item \textbf{Anode patterning:} The use of a gaseous medium for the conversion volume allows to use information of the ionization topology of the events as a handle to identify signals and reject background. To effectively use it the readout needs to be patterned with rather high granularity. Microbulk planes naturally allow for highly granular patterning and indeed, CAST detectors were the first 2D Micromegas readouts. Fig.~\ref{fig:2d_readout} shows different pattern concepts used in the CAST detectors, always with a pitch below 500~$\mu$m.

    \begin{figure}[t!]
\begin{center}
\includegraphics[height=4cm]{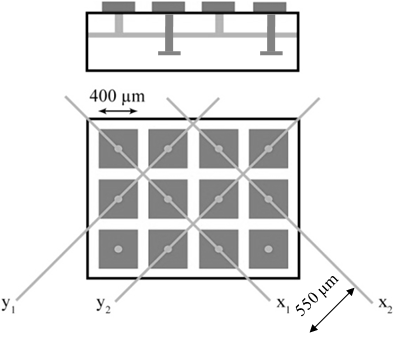}
\includegraphics[height=4cm]{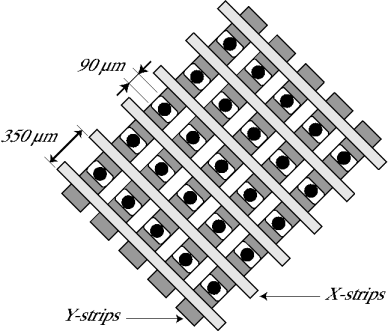}
\includegraphics[height=4cm]{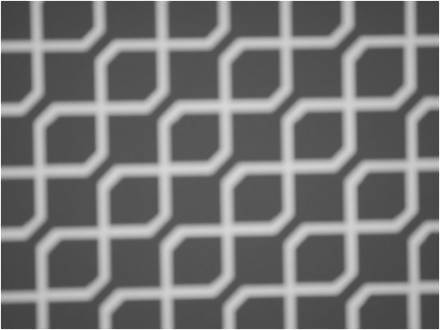}

\hspace{0.0cm}(a)\hspace{3cm} (b)\hspace{3cm} (c)
\caption{\label{fig:2d_readout} (a) Sketch of the 2-D readout strategy used in CAST; (b) shows the pattern of strips and pads used in the first detectors and (c) the most recent pattern based on interconnected pads. }
\end{center}
\end{figure}

\hyphenation{ma-nu-fac-tu-ring}
\item \textbf{Fabrication technology:} As mentioned before, microbulk Micromegas enjoy several improvements over more conventional techniques. Since their first application in CAST, the manufacturing process has been refined and consolidated.
\item \textbf{Radiopurity:} The raw material of microbulk readouts (i.e., double copper-clad kapton foils) as well as fully built readouts have been measured in underground Ge detectors~\cite{Cebrian:2010ta} and proven to be very radiopure materials. Moreover, a continuous effort have been made to study the radioactivity of other components of the detector body (chamber, x-ray window, screws, gas gaskets, connectors, etc.) and replace them with radiopure versions.
\item \textbf{Shielding:} Increasingly powerful passive and active shielding setups have been designed to reduce the contribution of external gammas to the detector background. Although shielding concepts from underground experimentation can be borrowed, care must be paid to the specifics of our case, e.g., the space and weight constraints of the moving platform, the operation at surface (presence of cosmic rays), the geometry imposed by the magnet (the shielding will always have an opening from which the signal x-rays reach the detector) and the intrinsic sensitivity and rejection capability of the Micromegas detectors. The most recent version of the CAST Micromegas shielding, based on a 10~cm thick pure lead shielding around a core of electroformed copper, is shown (partially built) in figure~\ref{fig:shieldinginstallation}.
\item \textbf{Offline rejection algorithms:} The detailed information obtained by the patterned anode, complemented with the digitized temporal wave-form of the mesh, is the basis to develop complex algorithms to discriminate signal x-ray events from other type of events. The power of this discrimination is highly coupled to the quality of the readout so that improvements in readout design or manufacturing yield improvements in discrimination power. The raw background in the energy region of interest (1 to 10 keV) is normally reduced by a factor of about 10$^2$ -- 10$^3$ only by offline discrimination.
\end{itemize}

As indicated in Fig.~\ref{fig:background_evol}, the lowest background achieved by Micromegas detectors in CAST is of below $10^{-6}$~\ckcs (sunset Micromegas detectors, 2013 data taking Run)~\cite{Aune:2013pna,mpgd2013_xavi}. In special setups underground, levels of $\sim10^{-7}$~\ckcs have been achieved~\cite{Aune:2013pna}. For IAXO we aim at background levels of at least $\sim10^{-7}$~\ckcs and if possible down to $\sim10^{-8}$~\ckcs (lower levels do not translate into better sensitivity as we reach the zero background situation for the exposure considered for IAXO). Below we describe the status of our knowledge of the background limiting factors, and the prospect to further improvement to reach the levels required by IAXO.

%

\subsection{Main sources of background: current understanding}

\begin{figure}[tb!]
\begin{center}
\includegraphics[height=17pc]{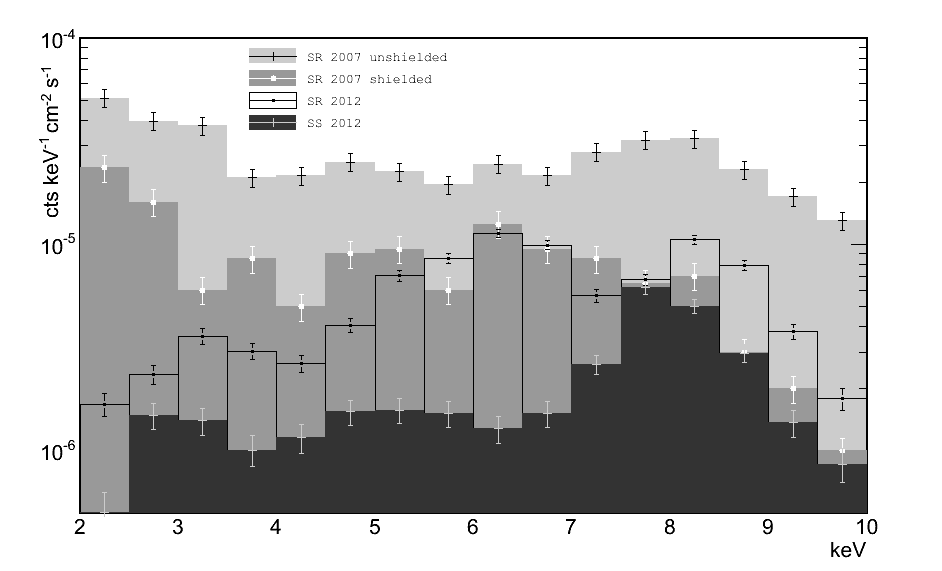}
\caption{\label{fig:background_evol}Evolution of the CAST micromegas background spectra since 2007. The lowest spectrum corresponds to the
2012 sunset data after the last improvement in shielding and the implementation of the cosmic muon veto.  }
\end{center}
\end{figure}

The improvements obtained are the result of the progressive understanding of background sources and their eventual rejection. This understanding has been achieved by a vigorous program of experimental tests and detailed simulations. Given that the final background level in a Micromegas detector is dependent on the application of the offline cuts on a number of event features, one needs complex simulations to reproduce the final background level from a given external radiation source~\cite{Aune:2013pna}. These simulations include not only Geant4-like simulations of particle transport and interaction through the detector setup, but also the full response of the detector, from the generation of the primary ionization in the conversion volume, to the drifting and diffusion of the electrons, amplification and induction of electronic signals in the electrodes. The level of reproducibility of the full detector response is currently very remarkable~\cite{ATomas_PhD}, and has been validated by means of calibrations and specific tests with gamma sources. Simulated data are generated with the very same format as real data from the DAQ, and they undergo the same analysis chain as the latter.

\begin{figure}[t!]
\begin{center}
\includegraphics[height=5.5cm]{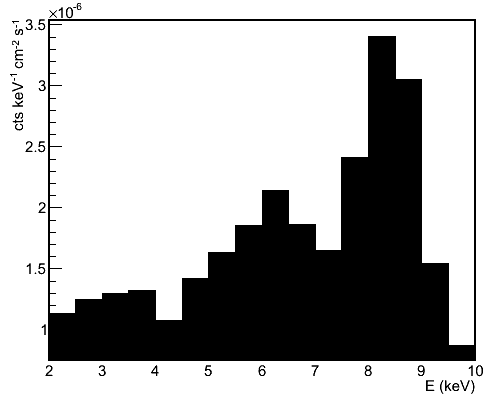}
\includegraphics[height=5.5cm]{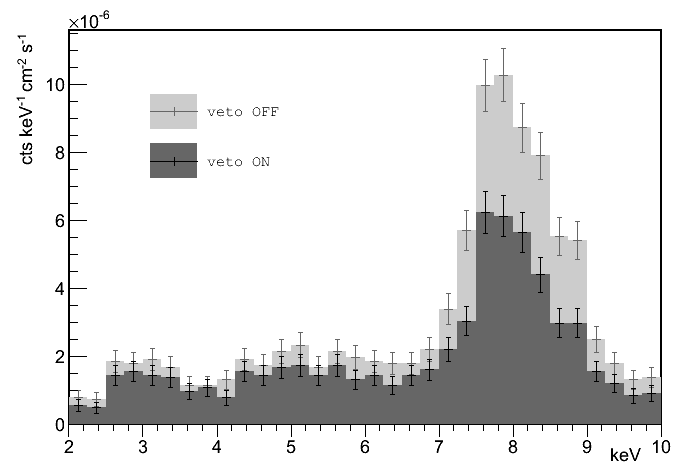}
\caption{\label{fig:gamma_flux}Left: simulated background spectrum induced by external $\gamma$ flux.
Right: background spectra in the CAST sunset Micromegas detectors in the 2012 Run, with and without the anticoincidence with the muon veto. About 25\% of the background in the 2-7 keV window are tagged as cosmic-ray-induced. Given the modest coverage of this veto, the result is compatible with the assumption that most of the current background is dominated by cosmics.}
\end{center}
\end{figure}

The background of the first generations of CAST Micromegas detectors was dominated by the external gamma radiation, and therefore most early efforts were directed towards quantifying and reducing it. Fig.~\ref{fig:background_evol} shows several experimental spectra obtained in different CAST setups from 2007 until 2012. In the next-to-last detector setup (labelled SR2012 in Fig.~\ref{fig:background_evol}), and using the simulation tooling developed, the external gamma radiation was quantified to contribute to about $\sim1.5\times10^{-6}$~\ckcs in the 2-7 keV range, the actual simulated spectrum being shown on the left of Fig.~\ref{fig:gamma_flux}. Although many different uncertainties affect these type of simulations (e.g., the precise spectrum and distribution of the simulated external gamma radiation) we consider this estimate accurate within a factor of 2-3. The shape of the spectrum also gives useful information, as it is populated by fluorescence peaks, produced by the remaining gammas reaching the innermost copper parts of the detector, or by gammas interacting in the stainless steel pipe that connects the detector to the magnet. This estimate is corroborated by experimental tests performed in CAST, as well as in test benches reproducing the same setups, both at surface and underground~\cite{Aune:2013pna}. This insight was used to design the last version of the shielding (labeled SS2012 in Fig.~\ref{fig:background_evol}), that should have reduced this contribution by, at least, one order of magnitude. The actual improvement, as shown in Fig.~\ref{fig:background_evol}, was more modest, due to the fact that cosmic rays have become the dominant source of background~\cite{Aune:2013pna}.

The level of background obtained by CAST 2012 detectors was $2\times10^{-6}$~\ckcs. According to our current best knowledge, this level is dominated by cosmic rays (muons) that cross the shielding and produce secondaries (mostly fluorescences) in the innermost parts of the setup, populating the low energy part of the spectrum. This assumption is sustained by tests done with partial-coverage muon veto working in anticoincidence (see Fig.~\ref{fig:gamma_flux}), as well as by the preliminary results of the CAST 2013 sunset detectors, which after an improvement of the active shielding show an improved background level of $\sim 8 \times 10^{-7}$~\ckcs~\cite{mpgd2013_xavi} (not included in Fig.~\ref{fig:background_evol}). A well-designed high-coverage active veto should be able to reduce this component of the background by one or two orders of magnitude. Efforts in this direction are planned for the near future as explained below in \ref{sec:prototype}.

Finally, another component of the background is the one induced by radioactivity from the detector components themselves. The current detector setup has undergone several redesigns and replacements of components regarding radiopurity (window strongback of copper instead of aluminium, gas inlets of copper instead of brass, etc.), and we estimate that the remaining internal radioactivity contributes to the detector background to, at most, $\sim 10^{-7}$~\ckcs. This level has been experimentally obtained in special shielding setups at LSC, in which both gamma (by full 4$\pi$ shielding) and cosmics (by operation underground) are reduced to negligible levels.

In summary, while current background levels at CAST are at the $\sim 10^{-6}$~\ckcs level, reduction to close to $\sim 10^{-7}$~\ckcs level seems at hand by an adequate design of an active veto. Further reduction below that level seems also realistic in the mid-term as a result of the ongoing campaign to identify and replace radioactive components of the detector. At lower levels of background we cannot exclude unidentified contributions from gammas entering through the remaining openings of the shielding facing the magnet, but they could be rejected by subsequent improvements in the shielding design.

\subsection{A demonstrating prototype of ultra-low background Micromegas detector for IAXO}
\label{sec:prototype}

As part of a coordinated effort to test both the optics and detector technologies proposed for IAXO, we are currently building a new version of Micromegas detector aiming at a further improvement in background level. This detector will be designed to operate at the focal point with a small scale x-ray optics. The combined set will be installed at the CAST sunrise side and will take physics data during 2014, giving precious operational experience on this combination of technologies for IAXO.

The new detector will be based on the current design, but will incorporate improvements from the results and considerations exposed before. The detector aims at reducing the background level down to, possibly, $\sim10^{-7}$~\ckcs. As argued previously, this level is feasible by improving the active and passive shielding. The main features of the new design are the following:


\begin{itemize}
\item \textbf{Shielding:} The new shielding will extend the concepts successfully tested in the 2012 CAST sunset setup. It will enclose better the detector geometry further reducing the opening solid angle in the direction of the magnet (taking advantage of the focusing of the optics). Active shielding with a muon veto will also be used, and a teflon coating will block fluorescence from the vacuum pipe.
\item \textbf{Detector chamber:} Shown on the right of Fig.~\ref{fig:det_concept}, the new chamber is mostly made out of electroformed copper, improving the radiopurity and acting as the innermost part of the shielding. The Micromegas readout have been redesigned with improved solder-less connections of all the drift and field cage electrodes. All front-end electronics components, potentially radioactive, are outside the shielding. All gaskets are made out of radiopure teflon.
\item \textbf{Electronics:} The new detector will enjoy an electronics based on the AFTER chip~\cite{Baron:2008zza} developed at CEA/Saclay to be used in the DAQ of the Micxromegas TPC of the T2K experiment. This electronics provides independent temporal information for every channel, adding extra topological information with respect to the previous (2D-projected) Gassiplex-based DAQ, and potentially adding discrimination capabilities to the offline analysis. The form factor of the electronics allows for a future upgrade to the AGET chip~\cite{Baron:2011} currently under development. This chip offers self-trigger capabilities which translate to potentially lower threshold for this detector.
%
%
%
\end{itemize}


The construction of this prototype, and its operation in 2014 in CAST in conjunction with the new x-ray optics will be an important milestone for IAXO. The detector is supposed to achieve a background level that will be very close to the requirement for IAXO. Its operation will give important information to define an additional decrease of the background well below $10^{-7}$~\ckcs. It is important to stress that the strategies for background reduction are not exhausted with the design here presented, and prospects for additional improvement are based on one or more of the following lines of work that will be studied in parallel:

%
%
\begin{itemize}
\item
Improvements in gamma shielding, especially towards further reduction of the unavoidable open solid angle in the connection of the detector to the magnet bore.
\item
Improvements in active muon shielding, increasing the veto efficiency.
\item
Improvements in radiopurity of components of the detector chamber itself.
\item
Improvements of the offline discrimination algorithm, possibly using the new topological information provided by the new AFTER electronics, and the lower threshold expected from the better signal-to-noise ratio offered by the new electronics.
In addition, further study of the discrimination algorithms will be possible with the dedicated multi--energy calibrations soon to be available in the x-ray tube installation of the CAST detector lab at CERN. Finally the use of different gases may bring improvement also in this direction.

\end{itemize}

\section{Additional infrastructure}
\label{sec:additional}

\subsection{General assembly, rotating platform and gas system}
\label{sec:platform}

\begin{figure}[b!]
\begin{center}
\includegraphics[width=7.5cm]{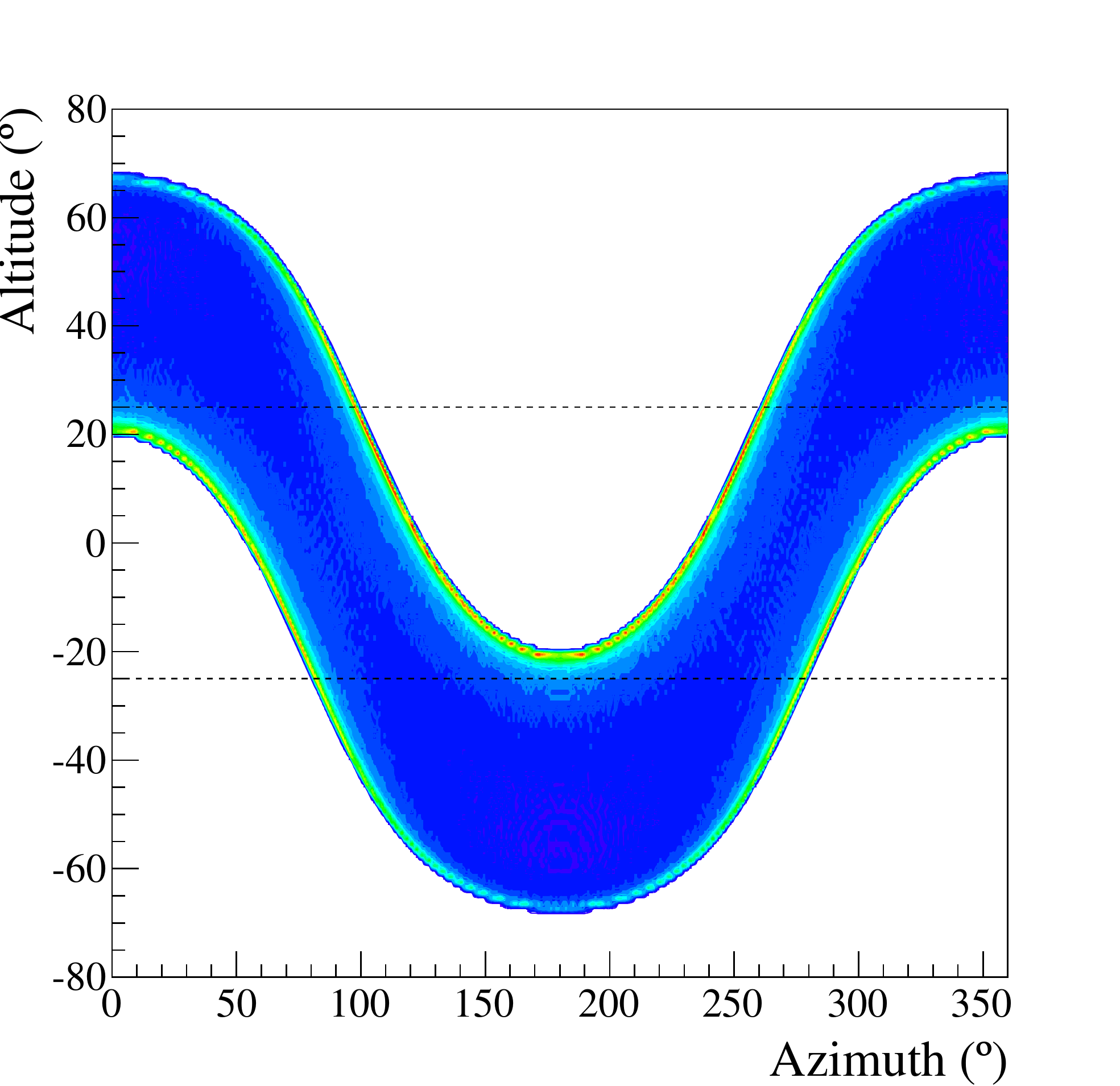}
\includegraphics[width=7.5cm]{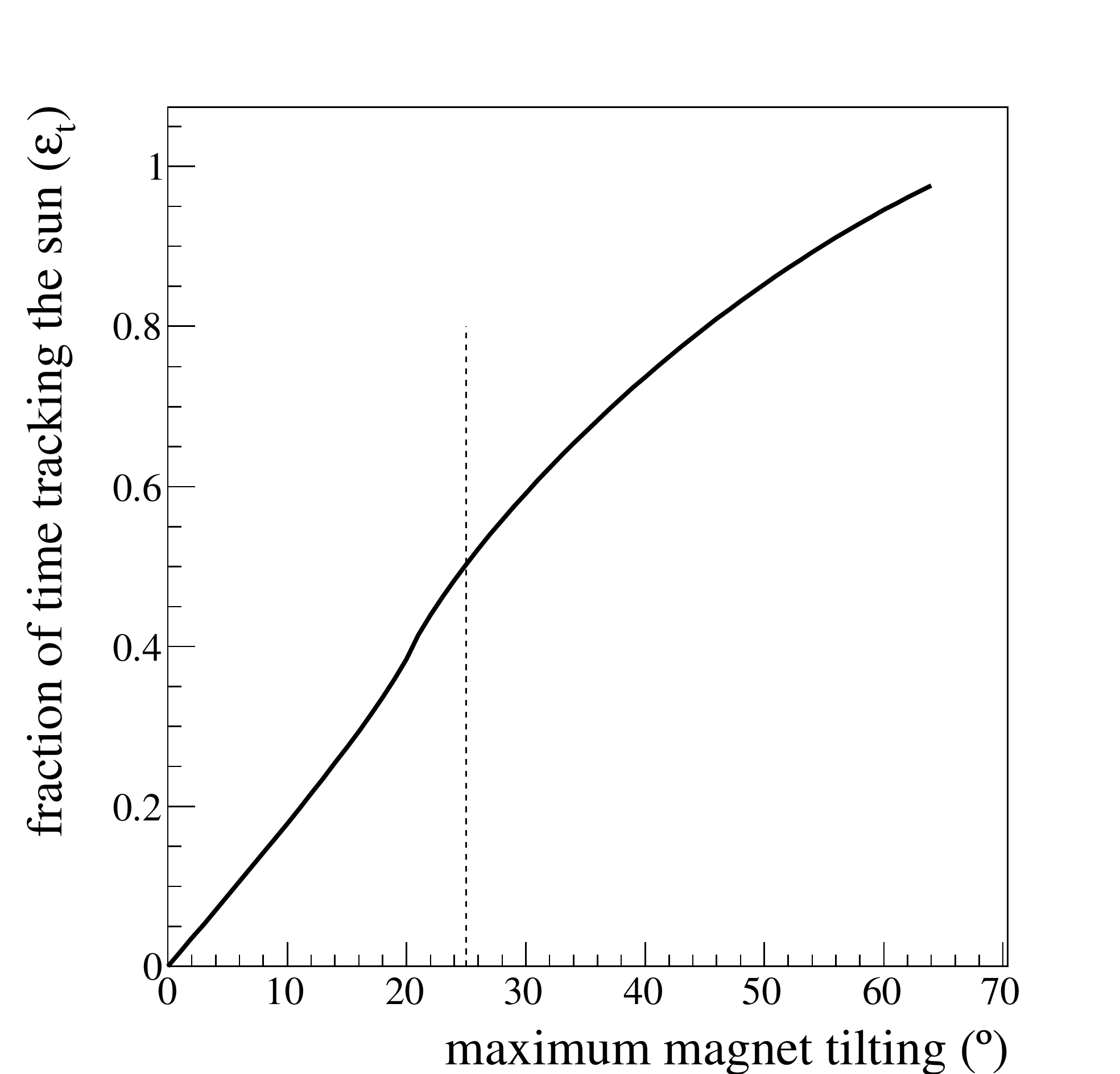}
\caption{\label{epsilon_t} On the left, the region of azimuth-altitude coordinates of the Sun position along one year for CERN geographical coordinates. The horizonal lines show the range of elevation for IAXO. On the right, the fraction of time the Sun is reachable versus the maximum elevation. The horizontal line indicates the value of elevation that we aim for IAXO, $\pm25^\circ$, which corresponds to fraction of time tracking the Sun $\epsilon_t =0.5$.}
\label{fig:histoPlot}
\end{center}
\end{figure}


The proper rotation and inclination of the IAXO magnet and detectors, which have to follow the sun's trajectory, is performed by an altitude-over-azimuth mount situated on a concrete foundation (see Fig.~\ref{fig:IAXO_sketch}). This assembly is generating the system's motion via two semi-circular objects, responsible for the inclination by means of hydraulic pads and rollers, and a structural steel disk, mounted on circular rails and generating the system's rotation by means of roller drives. The primary magnet system service units, including the interfacing proximity cryogenics enforcing supercritical helium flow in the cold mass cooling pipes, the cryostat isolation vacuum system, the power convertor and auxiliary switches, current leads and the diodes-dump-resistors unit, as well as the controls and safety systems are mounted on top of the rotating table to couple their position to the rotation of the magnet cryostat, thereby simplifying the transfer lines and flexible cables arrangement. As presented in section~\ref{sec:magnet} and shown in Fig.~\ref{fig:IAXO_sketch}, all service lines connected to the magnet cryostat are provided through flexible transfer lines and flexible superconducting cables that are connected to a stationary services turret on the magnet cryostat. The flexibility of the transfer lines and cables is required to compensate for the changing inclination angle of the magnet. The ground stationary service units, like the cryogenics plant providing liquid helium to the system and mains switchboards, will be housed in the side building.

\begin{figure}[t!]
\begin{center}
\includegraphics[height=9cm]{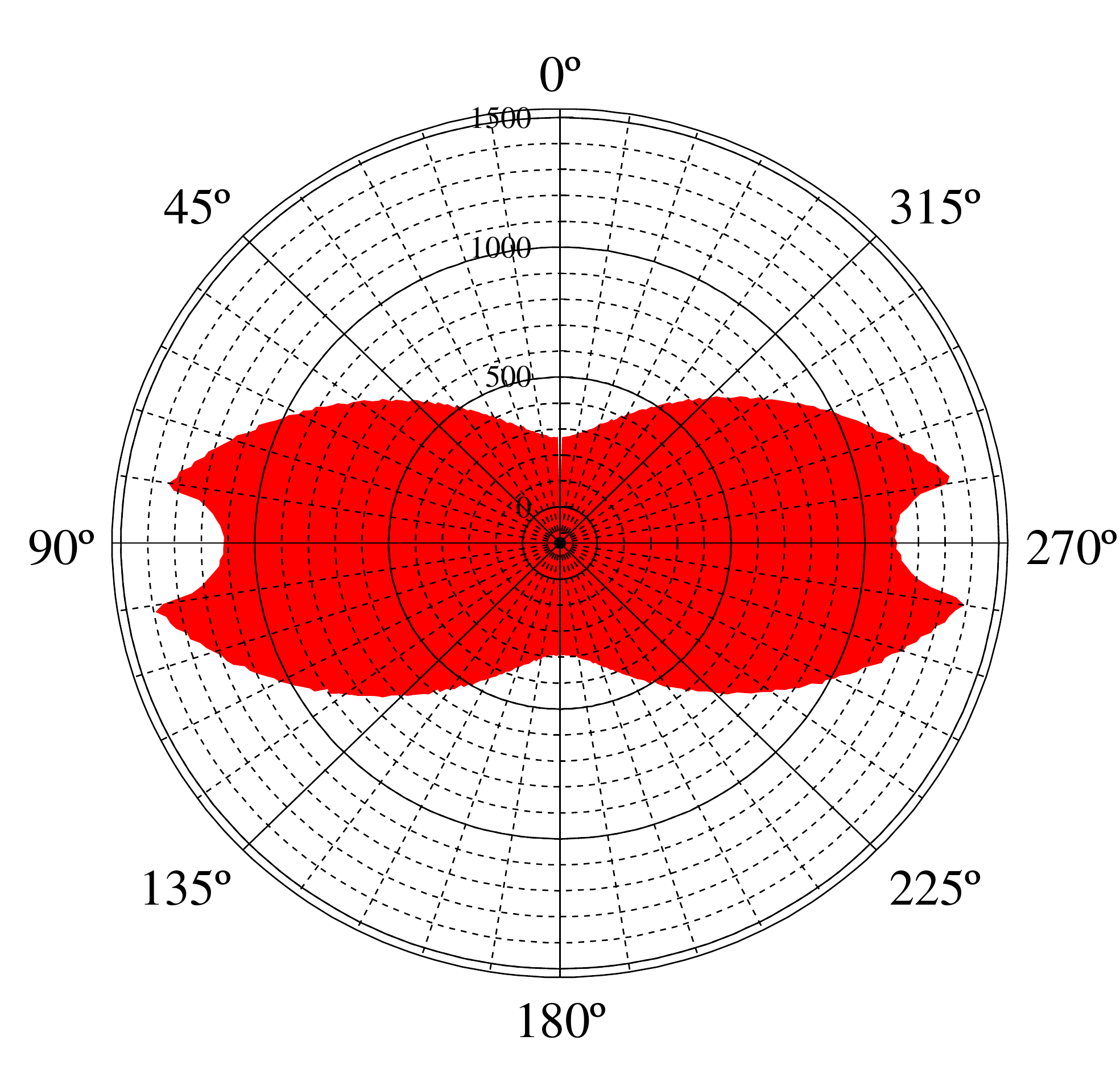}
\caption{\label{fig:azimuthdist} The distribution of exposure time in azimuthal coordinates. North direction corresponds to 0$^\circ$}
\label{fig:polar}
\end{center}
\end{figure}

The movement of the Sun in the azimuth-altitude plot along the year is shown on the left of Fig.~\ref{fig:histoPlot} for CERN geographical coordinates. Assuming all the azimuthal angles are reachable, the maximum elevation of the structure will determine the total fraction of time that the sun is reachable by IAXO. On the right of Fig.~\ref{fig:histoPlot} the relation between maximum magnet tilting and the fraction of time of sun tracking is shown, computed for the CERN geographical coordinates. If we aim at an effective exposure of about half of the time ($\epsilon_t =0.5$), the needed maximum elevation for IAXO's structure is $\pm25^\circ$. In Fig.~\ref{fig:polar} we show, in polar coordinates, the distribution of exposure time to the Sun in azimuthal coordinates.

In addition, for the second data taking campaign (what we call Run II in~\cite{Irastorza:1567109}) the magnet bores must be filled with gas at precise and well-controlled pressures, for which an appropriate gas system will be needed. The system will have several stages including storage, circulation, pumping, metering system and a set of monitoring probes for the gas. The experience with the gas system developed for CAST will be very valuable as conceptually the system will be similar, however, the system needed for IAXO will be substantially simpler in several aspects. For CAST the operating gas was $^4$He and at a later stage $^3$He (a considerably expensive gas that required very stringent safety requirements to the system). $^3$He was needed to reach the high densities associated to the values of $m_a$ aimed for, given that the gas was at cryogenic temperatures when inside the bore. The fact that IAXO bores are at room temperature and that the maximum axion mass aimed for is of $m_a \sim 0.25$~eV~\cite{Irastorza:1567109}, makes that the IAXO gas system will use only $^4$He, the gas will be always at room temperature and the working pressures will range from 0 to 1 bar (no need for high pressure specifications). In CAST conditions, the need to study and control the potential excursions and inhomogeneities of the gas density along the magnet bore implied a large challenge for the monitoring, simulation and data treatment of the experiment. Although detailed studies of, e.g., the expected density gradient along the IAXO bores will be done for the Technical Design Report, the maximum densities of the gas in IAXO will always be well below the densities where we expect any substantial effect in density gradient according to the experience in CAST.

\subsection{Additional equipment}
\label{sec:additional}

The equipment described in previous sections completes the baseline of IAXO. Magnet, optics and x-ray detectors have been described to some extent and rely on solid technological ground. They are sufficient to accomplish the primary physics goal of the experiment. The additional equipment described below rely on less developed considerations but offer potential improvements beyond the baseline of the experiment. They include GridPix detectors, Transition Edge Sensors (TES), low noise Charge Coupled Devices (CCD) and microwave cavities and antennas. None of the quantified physics potential explained in~\cite{Irastorza:1567109} relies on any of these devices. The motivations to consider an complementary or alternative use of this equipment are several:

\begin{itemize}
\item They offer potential to span the detection energy window of solar ALPs to lower energy ranges (GridPix, TES, CCD) which could be interesting for additional searches of more specific ALPs or WISP models (e.g., hidden photons, chamaleons or other ALPs)
\item They offer potential for new physics cases, like the detection of dark matter axions and ALPs (microwave cavities or antennas).
\item Even if at the moment the Micromegas detectors are the ones with best prospects to achieve the needed FOM for low background x-ray detection in IAXO, other technologies, e.g., low noise CCD, could eventually prove competitive too as R\&D is ongoing in this direction. Provided similar FOMs were achieved by a second technology, the preferred configuration for IAXO would be a combination of the two (e.g. half the magnet bores equipped with one kind of detector and half with another), as this configuration is more immune to systematics effects in case of a positive detection.
\item Finally, the consideration of more technologies is helpful to attract and built community around the project.
\end{itemize}

\subsubsection{GridPix}

\begin{figure}[b!]
\begin{minipage}{0.48\linewidth}
\centering
\includegraphics[width=1.0\linewidth]{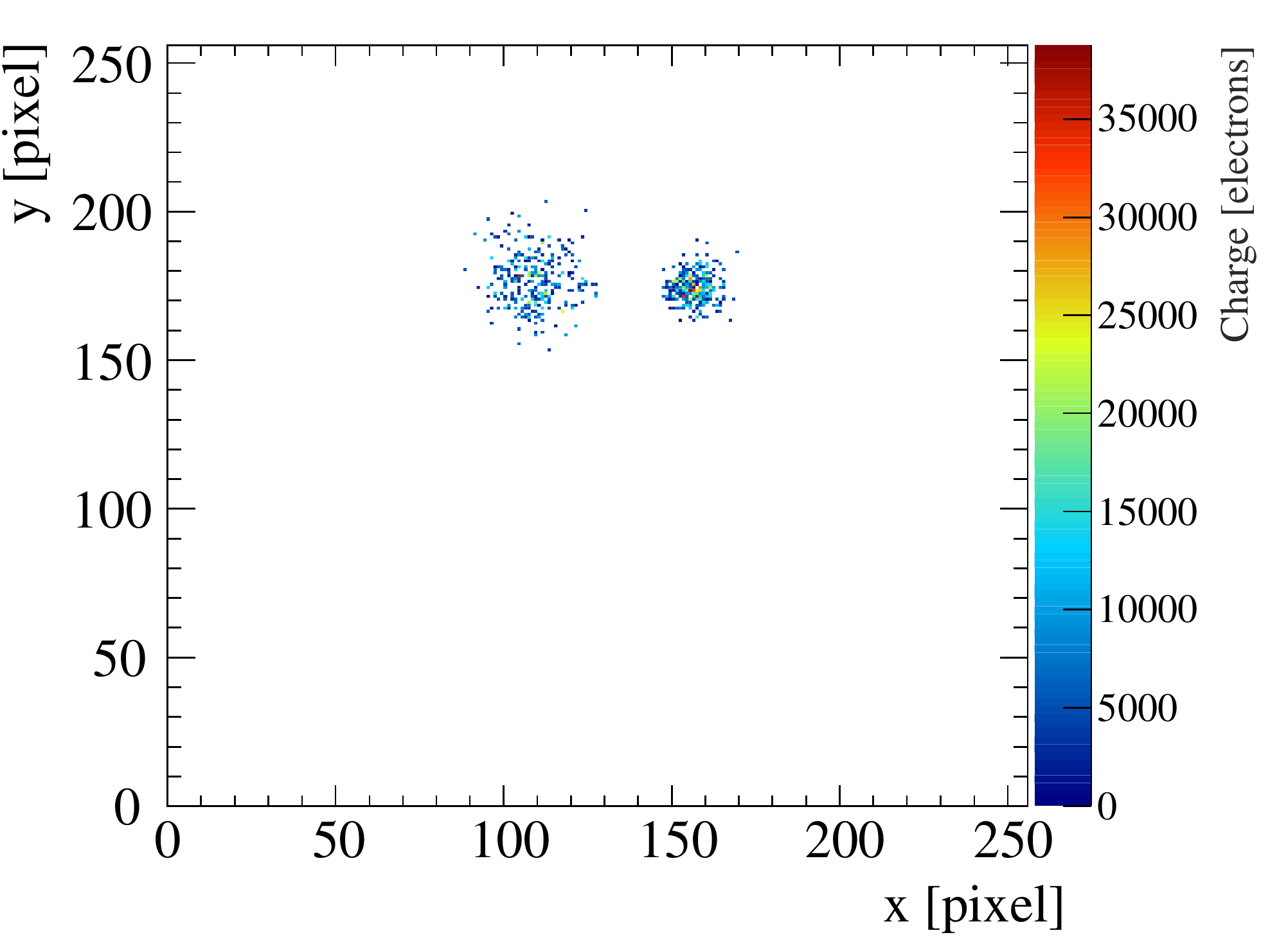}
\end{minipage}
\begin{minipage}{0.48\linewidth}
\centering
\includegraphics[width=1.0\linewidth]{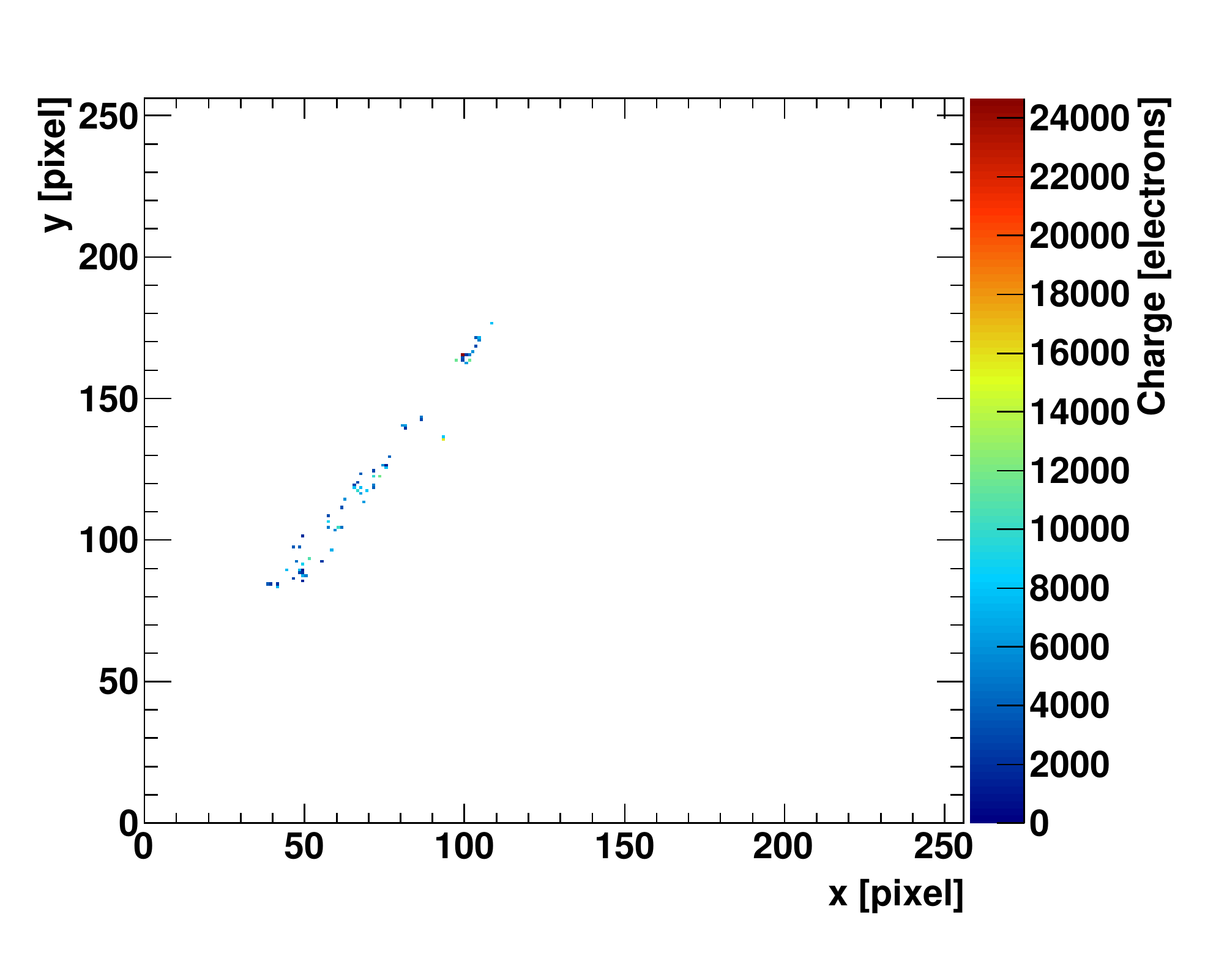}
\end{minipage}
\caption{\label{fig_Fe55} Left: Conversion of two X-ray photons at different
  drift distances, right: track of cosmic ray~\cite{proc_ingrid}.}
\end{figure}

The GridPix development aims at the combination of a Micromegas with a
highly pixelized readout~\cite{Chefdeville:2005ku}. To cope with the large number of
electronics channels needed to cover significant areas with very fine granularity, the readout ASICs of pixel detectors are
used. The bump bond pads, which are usually used to connect the ASIC to the silicon sensor, are placed below the Micromegas and serve as charge
 collection pads. Currently the Timepix ASIC~\cite{Llopart2007485} is used
 as a readout chip, but the successor, Timepix3, is under development
 and will feature many improvements of the Timepix's limitations. The
 Micromegas gas amplification structure can be built directly on top of the
 ASIC. Industrial post-processing techniques ensure an optimal alignment of
 grid holes with the pixels, a very thin grid of less than 1~$\mu$m
 thickness and a very homogenous gap and hole size. The low noise level and
 uniform gas amplification properties of the setup result in a single
 electron efficiency of 100 \% for moderate gas amplifications of a few
 thousand (the event picture with 2 photon conversions of an $^{55}$Fe-source are shown on the left of Fig. \ref{fig_Fe55}). If sufficient diffusion is provided, the primary electrons of each photon conversion can be determined by counting the pixels hit.

 This results in three features potentially very interesting for IAXO. Firstly, it should lead to improved energy
 resolution (see for example Ref. \cite{unser_cast_paper}). Secondly, when a
 photon is absorbed in the gas, on average one electron is created
 per 20-30 eV photon energy. Only a few primary electrons are sufficient to
 identify a photon, thus lowering, in principle, the threshold for photon
 detection. Currently, first
 measurements are under way to study the lower energy part of the detectable
 spectrum. Finally, the precise topological reconstruction of events helps to
 distinguish the photon conversions from background events such as cosmic ray
 tracks (the event picture with of a background track can be seen on the right of Fig. \ref{fig_Fe55}). With a
 first detector a background suppression by a factor of about 120 was
 reached~\cite{unser_cast_paper}. With more sophisticated
 algorithms such as neural nets, and new electronic components, in particular
 the above-mentioned Timepix3, will lead to further improvements in all three
 key parameters. First tests of such a system are planned for the fall of 2013
 in the CAST experiment. The development of critical detector components, such
 as photon windows with low material budget, will continue independently.

\subsubsection{Transition edge sensors (TES) for IAXO}


Transition Edge Sensors (TES)~\cite{irwin2005} offer the possibility of extending the detection energy window of IAXO to much lower energies while potentially keeping single-photon counting capability and very low background rates. TES-based photon sensors
have an extremely low intrinsic dark count rate since they operate at sub-K temperatures, where the available thermal energy is not sufficient to excite a transition. The response time of TES-based sensors is in the range of the tens of microseconds, not sufficient for fast counting applications, but more than adequate in WISP searches, where expected counting rates are well below 1 Hz. Within these limitations TES-based sensors are capable of counting single photons down to the sub-eV energy range \cite{miller2003,lita2008,angloher2009}.
The energy range of maximum sensitivity for TES-based sensors depends on the thickness and type of material used for the layer which must absorb the incoming photons. It is possible to manufacture sensors with overlapping energy sensitivities from $\sim$1 eV up to several keV and cover all the energy ranges of interest of solar WISP detection: visible (hidden photons), below 1 keV (chameleons), keV range (QCD axions and other ALPs).

\paragraph{Challenges for TES operation at helioscopes}

TES-based sensors operate in a sub-K environment (typically 100 mK) which must be reached and maintained for the entire duration of the measurements. The standard way to operate a TES is to insert it into a refrigeration unit which uses cryogenic fluids or a pulse tube to bring the temperature in the K range, and then either $^4$He-$^3$He dilution or Adiabatic Demagnetization Refrigeration(ADR) to move down to the mK regime. All these cooling solutions are commercially available. In a magnetic helioscope, such as IAXO, the first-stage cooling needs (from room down to K temperatures) could be covered by using the liquid He supply necessary for magnet operation, while the second-stage cooling technique must be chosen between dilution and ADR. The former (dilution) has the advantage of working in continuous mode once the final temperature is reached, and of being largely insensitive to stray magnetic fields, while disadvantages include the necessity of precisely manipulating expensive gases such as $^3$He. The latter (ADR) has the advantage of ease of operation and the disadvantage of working in hour-long cycles: the cold chamber slowly heats up and the demagnetization procedure must be repeated. In addition, since the SQUID circuits necessary for TES readout are sensitive to the stray magnetic fields generated by the ADR, shielding and compensation coils must be accurately designed.

The main drawback for the use of TES-based sensors in a helioscope is the small sensitive area: typical sensors have sensitive areas of (100 $\mu$m)$\times$(100 $\mu$m) up to (1 mm)$\times$(1mm). This limitation can be partially overcome by combining single sensors into an array connected in parallel~\cite{dekorte2002,smith2009}. Such arrays could reach sensitive areas of about 1 cm$^2$. An additional possibility is to position the TES-based sensor in the focal point of an x-ray optics. The focal point area considered for IAXO optics are $\sim$0.2~cm$^2$. The reflectivity of the optics to the appropriate photon wavelengths should be taken into account.
 A further challenge is coupling the TES-based sensor to the helioscope beam. In the case of $\sim$1eV photons this can accomplished by means of an optical fiber conveying photons from the helioscope to the sensor itself which can, in principle, be positioned far away from the magnet and within its own refrigeration unit. With photons of energies above $\sim$10 eV, the TES-based sensor must be placed directly in the beam.

\paragraph{Proposed detector}

The basic characteristics of TES-based sensors are ideally suited for WISP searches, while the small sensitive area presents a challenge when proposing the application of these sensors to helioscopes such as IAXO. One option to turn this weakness into a strength is to combine several TES sensors sensitive to different energy ranges into a single chip to be positioned in the focal point of an x-ray optics. Such a Multy-Energy TES Array (META) could be for example one of the basic detectors to be used in IAXO runs.
Possible construction steps for META:

\begin{itemize}
\item Design and manufacture a single chip with TES sensors having different absorber thicknesses/materials in order to cover photon energies from 1 eV up to several keV. Each sensor must have its own current readout channel, keeping however the option of connecting them all in parallel. Target size for individual sensor sensitive area should be (1 mm)$\times$(1 mm).
\item Test multi-energy TES arrays in a stand-alone refrigerator (requires finding appropriate photon sources)
\item Design and manufacture cold finger assembly to hold the META sensor in the beamline after choosing a final-cooling stage method (either dilution or ADR)
\end{itemize}

Sensor pads differing only in thickness can be integrated on a single chip with readily available technology and equipment. For pads relying on different layer material acting as absorbers, some technical developments are necessary. Furthermore, TES pads having layers made of the same materials will work at the same transition temperature, while for pads made of different materials the question must be investigated through simulations and actual tests.
There is also the possibility to individually tune the electro-thermal feedback biasing each pad to compensate for differences in transition temperature and response, this must however be thoroughly checked experimentally. Finally, the technology to achieve a sensitive area of 1 mm$^2$ area for each sensor pad is presently available, however this development will require a significant increase in financial and personnel resources with respect to the present situation.

In practice, the possible short term developments are to design and manufacture a prototype META chip and to conduct laboratory tests with visible and soft X-ray sources. With readily available equipment, a prototype META could be manufactured with 2 or more (20 $\mu$m)$\times$(20 $\mu$m) TES pads made of the same material, but having different thicknesses in order to cover photon energies up to 1 keV.

It should be noted that the ALPS-II experiment at DESY in Hamburg is successfully operating a TES sensor in an ADR with very low backgrounds and an energy resolution below 10\% for single 1064 nm photons. It is planned to use this TES for WISP searches in 2015. Future space-born X-ray missions are developing TES arrays as detector options. The BaRBE collaboration in Italy (INFN Trieste, University of Camerino, INRiM Torino) is presently testing TES-based detectors sensitive in the visible range and is developing coupling methods with source photons using optical fibers. The intended application is WISP searches, including CAST and IAXO. Hence, some experience exists in the application of TES systems in the fields relevant for IAXO.

\subsubsection{Low-noise Charge Coupled Devices (CCDs) for IAXO}


Charge coupled devices (CCDs) are routinely used for detection of soft x-rays or photons with near-visible wavelengths, in applications like photography or astronomical imaging. These devices have been developed in part to make use of their small pixelated structures which allow for high resolution image reconstruction. In this standard application, noise is often not a problem as the integration time (exposure time) is relatively short and thus there is low noise even with fast readout. Image processing algorithms can be employed
to clean up residual artifacts from noise. As a static detector integrating energy deposited by
photons or other particles, CCDs have a noise component that arises from internal dark currents and the accumulation of charge deposited by
the clocking signals used to move charge to readout amplifiers within the CCD. Such noise makes CCDs less than ideal in the search for
signals from new particles as the differentiation between charge from noise and charge from possible new physics is difficult to distinguish.

R\&D on techniques to minimize this noise have been carried out with CCDs used for the Dark Energy Survey (DES)
astronomical survey. These CCDs, in order to be more sensitive to near infra-red photons, are hundreds of microns thick compared with more traditional
astronomical CCDs. An interesting by-product application found for these more massive CCD sensors is to use them as the sensitive element in the search for low-mass WIMP Dark Matter recoils. The DArk Matter In CCDs (DAMIC)~\cite{Barreto:2011zu}
experiment is using CCDs in the search for low-mass WIMPs where low noise is necessary in order to gain sensitivity
for their interactions in the bulk of the silicon.

There are plans to study the possibility of using this type of low noise CCDs in IAXO. This could be either as alternative keV x-ray detector, provided the ongoing low noise R\&D succeeds in getting a competitive figure of merit, or as a detector for softer x-rays, in order to expand the IAXO energy window to lower values, as needed for a number of additional potential physics goals of IAXO~\cite{Irastorza:1567109}. CCDs have been used as detectors for soft x-rays and thus there is nothing intrinsic in preventing the DES CCDs being used in a solar helioscope although the quantum efficiency versus x-ray wavelength would need to be quantified. Soft x-rays also do not penetrate much material and a suitable window or other system would be required to ensure that x-rays produced in the helioscope would impinge upon the sensor. As long as these technical difficulties are addressed, there could be a real advantage to using low noise CCDs in a helioscope to search for the new particles in a manner that would complement more conventional approaches to x-ray detection.

\subsubsection{Microwave cavities and/or antennas}
\label{sec:additional_mw}

Still at an early stage of development, the possibility of using the large magnetic volume of IAXO to search for relic axions or ALPs is very appealing~\cite{Irastorza:1567109} as these particles are excellent candidates for the Dark Matter of the universe. Searches for Dark Matter axions require microwave cavities or -as recently proposed- antennas, coupled with appropriate low noise sensors, all embedded in strong magnetic fields. So far, only one true axion Dark Matter experiment is running and a huge discovery potential could open up with IAXO infrastructure.
The sizes, design and features of these elements will depend very much on how this concept is further developed. First discussions have been stimulated and small experiments are on the way~\cite{Redondo:2013hca}.
The fact that the IAXO magnet is built with very easy access to the magnet bores make the option to instrument one or more of these bores with cavities or antennas feasible. This could be conceived as an additional data taking phase or in parallel with the solar axion searches. Technically more difficult, but also conceivable, would be to instrument the unused empty magnetic space inside the magnet cryostat for this purpose.

\section{Conclusions}
\label{sec:conclusions}

We have presented the conceptual design of IAXO, the International Axion Observatory, a proposed forth generation axion helioscope. IAXO follows the layout of an enhanced axion helioscope, and is based on a purpose-built 20-m long 8-coils toroidal superconducting magnet. All the eight 60-cm diameter magnet bores are equipped with focusing x-ray optics, able to focus the signal photons into $\sim$0.2~cm$^2$ spots that are imaged by ultra-low-background Micromegas x-ray detectors. The magnet is built into a structure with elevation and azimuth drives the will allow for solar tracking for ~12 h each day. All the enabling technologies exist, there is no need for development. All the needed know-how is present in the proponent groups. IAXO has been recently proposed to CERN~\cite{Irastorza:1567109}.

The anticipated experimental parameters of IAXO indicate a signal-to-background ratio, of about 4$-$5 orders of magnitude higher than CAST, which translates into a factor of $\sim20$ in terms of the axion-photon coupling constant $\gagamma$. That is, IAXO will deeply enter into completely unexplored ALP and axion parameter space. Potential additional physics cases for IAXO are the search of axionic dark radiation, relic cold dark matter axions or the realization of microwave light-shining-through wall setups, as well as the search of more specific models of weakly interacting sub-eV particles (WISPs) at the low energy frontier of particle physics. IAXO has the potential to serve as a multi-purpose facility for generic axion and ALP research in the next decade.

\section*{Acknowledgements}
\label{sec:acknowledgements}

We acknowledge support from the Spanish Ministry of Science and Innovation (MICINN) under contract FPA2008-03456 and FPA2011-24058, as well as under the CPAN project CSD2007-00042 from the Consolider-Ingenio2010 program of the MICINN. Part of these grants are funded by the European Regional Development
Fund (ERDF/FEDER). We also acknowledge support from the European Commission under the European Research Council T-REX Starting Grant ERC-2009-StG-240054 of the IDEAS program of the 7th EU Framework Program. Part of this work was performed under the auspices of the U.S. Department of Energy by Lawrence Livermore National Laboratory under Contract DE-AC52-07NA27344 with support from the LDRD program through grant 10-SI-015. We also acknowledge support from the CAST collaboration. The design work on the magnet system was supported by CERN, Physics Department as well as the ATLAS Collaboration. Partial support by the Deutsche Forschungsgemeinschaft (Germany) under grant EXC-153, by the MSES of Croatia and the Russian Foundation for Basic Research (RFBR) is also acknowledged. F. I. acknowledges the support from the Eurotalents program.

\footnotesize
\bibliographystyle{JHEP}
\bibliography{../../../BibTeX/igorbib,redondobib,axion_LOI_optics}

\end{document}